\documentclass[a4paper,8pt, english]{article} 

\usepackage{amsmath,amssymb,amsfonts}
\usepackage{algorithmic}
\usepackage{graphicx}
\usepackage{textcomp}
\usepackage{xcolor}
\usepackage{booktabs} 
\usepackage{gensymb}
\usepackage{array}
\usepackage{tcolorbox}
\usepackage{hyperref}
\usepackage[margin=0.8in]{geometry}
\usepackage{lscape}
\usepackage{adjustbox}
\usepackage{indentfirst}
\usepackage{multicol, blindtext}
\usepackage[font=small]{caption}

\usepackage{ragged2e}

\hypersetup{%
colorlinks=true,
linkcolor=blue,
citecolor=blue,
urlcolor=blue}

\usepackage{verbatim}

\begin{document}

\centering
\textbf{\LARGE{Assessing the embodied carbon footprint of IoT edge devices with a bottom-up life-cycle approach}}\\
\large{Thibault Pirson$^{a,*}$ and David Bol$^a$} \\

\flushleft
\small{$^a$\textit{Université catholique de Louvain, ICTEAM/ECS, Louvain-la-Neuve, Belgium}} \\
\small{$^*$\textit{Corresponding author at:} Bâtiment Maxwell a.193, Place du Levant, 3 (L5.03.02) 1348 Louvain-la-Neuve, Belgium.} \\
\small{\textit{Email address:} thibault.pirson@uclouvain.be}\\
\small{\textbf{Submitted for review}}\\

\centerline{\rule{17cm}{0.8pt}}
\vspace{-0.2cm}
\justify
\textbf{Abstract}
In upcoming years, the number of Internet-of-Things (IoT) devices is expected to surge up to tens of billions of physical objects. However, while the IoT is often presented as a promising solution to tackle environmental challenges, the direct environmental impacts generated over the life cycle of the physical devices are usually overlooked. It is implicitly assumed that their environmental burden is negligible compared to the positive impacts they can generate. Yet, because IoT calls for a massive deployment of electronic devices in the environment, life-cycle assessment (LCA) is an essential tool to leverage in order to avoid further increasing the stress on our planet. Some general LCA methodologies for IoT already exist but quantitative results are still scarce. In this paper, we present a parametric framework based on hardware profiles to evaluate the cradle-to-gate carbon footprint of IoT edge devices. We exploit our framework in three ways. First, we apply it on four use cases to evaluate their respective production carbon footprint. Then, we show that the heterogeneity inherent to IoT edge devices must be considered as the production carbon footprint between simple and complex devices can vary by a factor of more than $150 \times$. Finally, we estimate the absolute carbon footprint induced by the worldwide production of IoT edge devices through a macroscopic analysis over a 10-year period. Results range from 22 to 562 MtCO2-eq/year in 2027 depending on the deployment scenarios. However, the truncation error acknowledged for LCA bottom-up approaches usually lead to an undershoot of the environmental impacts. We compared the results of our use cases with the few reports available from Google and Apple, which suggest that our estimates could be revised upwards by a factor around $2\times$ to compensate for the truncation error. Worst-case scenarios in 2027 would therefore reach more than 1000 MtCO2-eq/year. This truly stresses the necessity to consider environmental constraints when designing and deploying IoT edge devices. 

\centerline{\rule{17cm}{0.8pt}}
\vspace{-0.2cm}
\justify

\section{Introduction} \label{sec:intro}

The Internet-of-Things (IoT) aims at connecting every physical object to the cloud, everywhere, all the time. The number of IoT edge devices which will be deployed is difficult to estimate and will vary from 5 billion in 2020 up to 200+ billion in 2030 \cite{Statista_IoT, Gartner_2018, CISCO, sparks2017route_ARM, strous2019internet}. Nevertheless, a rapid increase of their number has been acknowledged \cite{CISCO} and is widely predicted for the upcoming years. Electronic components will be incorporated into everyday appliances, transforming them into more sophisticated systems connected over the Internet and mobile networks \cite{ITU_WEEE}. Indeed, such connected devices require computing capabilities and connectivity features to transmit data acquired within their environment from the edge to a centralized management system. \\

IoT is part of Information and Communication Technologies (ICT), which are frequently presented as a very promising way to move towards less resource-intensive societies. It relies on their ability to monitor and optimize complex systems in real time, referred to as positive enabling effects \cite{hilty2015ict, berkhout2001impacts}. Fields of application range from smart cities to smart health, smart agriculture and smart mobility, to name only a few. \\

However, ICT also contributes increasing environmental impacts throughout the life cycle of its services and products \cite{plepys2002grey}. These impacts are referred to as direct impacts \cite{hilty2015ict, berkhout2001impacts} and occur during material extraction and refining, manufacturing, transport, use and disposal. Several papers agree on the current absolute ICT carbon footprint which is evaluated at about 1000-2000 MtCO2-eq or equivalently 2-4$\%$ of the world greenhouse gas (GHG) emissions \cite{freitag2021climate}. However, there is no consensus on future trends. As IoT is a growing part of ICT, it should also be subject to analyses regarding environmental impacts and sustainability \cite{IoTSustBook}. This is supported by existing LCA of ICT devices \cite{ercan2016} showing that the production of electronic components accounts for a major part of the environmental impacts, especially for battery-powered devices. However, very few LCA results are available for IoT edge devices, thus preventing an objective comparison between direct burdens and positive enabling effects \cite{Gray2018, quisbert2020lifecycle, hittinger2019}. Furthermore, the task is even more difficult as IoT gathers heterogeneous designs for many different applications, which challenges systematic LCA studies and comparisons. \\

In this paper, we propose a framework based on hardware profiles to streamline the carbon footprint evaluation of IoT edge devices production. It focuses on consumer IoT edge devices which have been pointed out as the main contributor to the rising number of IoT devices \cite{CISCO}. The framework can also serve as a basis for Industrial IoT devices, Industry 4.0 applications and other smart related applications. Quantitative results are given for four IoT use cases and we demonstrate the vast range of carbon footprints that can be attributed to IoT edge devices, up to a factor $158\times$. This clearly highlights the large inaccuracy that can arise if the heterogeneous nature of IoT is not considered. We then use the framework to conduct a macroscopic analysis in order to evaluate the carbon footprint of the massive production of IoT edge devices over a 10-year period (2018-2028). This reveals a problematic carbon footprint for massive IoT production if IoT edge devices consist primarily of complex devices.\\

This paper is organized as follows: Section \ref{sec:sota} gives a literature review of environmental impact assessment of IoT. Section \ref{sec:methodology} describes the methodology used to build the framework based on hardware profiles. It is then applied to four IoT edge devices and quantitative results are presented in Section \ref{sec:results}, together with an interpretation of the results, a comparison with existing studies, and a sensitivity analysis. Finally, the environmental impacts of massive production of IoT edge devices are discussed in Section \ref{sec:discussion}.
\section{Background} \label{sec:sota}

There is a considerable literature analyzing the enabling effects of the IoT and motivating the transition towards a more efficient use of resources, a better monitoring of the supply chains, and the development of a circular economy \cite{ingemarsdotter2019circular, van2018footprint, mishra2019carbon, mois2017analysis, tao2014internet}. However, far less research has been conducted on the direct impacts generated by the deployment of the IoT. This section focuses on the latter by presenting the main contributions in the literature. It reveals that while general methodologies already exist, quantitative results with sufficient details on hardware and assumptions considered are still scarce for assessing the direct environmental impacts of IoT.  \\

Generic IoT device life-cycle models are proposed in \cite{soos2018iot, rahman2018} but even though they are helpful for a better understanding of the different key steps in the life-cycle of an IoT device, they do not provide quantitative results to model the direct impacts of IoT devices.  \\

Several studies attempt to consider the burden generated by the IoT devices in specific applications but they usually use over-simplified modeling mainly because of the high complexity of electronic devices and the lack of data. The low resolution on electronic components is therefore a bottleneck. The authors of \cite{van2013home, van2013smart_book} have questioned the environmental consistency of multi-functional home energy management system (HEMS) by including the smart infrastructure burden into the balance. However, the assessment was done by scaling a reference common product i.e. printed circuit board (PCB) for a laptop mainboard, which is most likely not representative of the actual HEMS hardware. Home sensing and automation systems have also raised concerns in \cite{bates2013exploring}. They emphasize the importance of accounting for additional life-cycle impacts arising from the introduction of the connected layer itself and not only all the potential savings generated. Quantitative results are given for the embodied carbon footprint of several sensors but the lack of details and the publication date (2013) limit the re-usability. A recent paper \cite{babbitt2020disassembly_nature} offers a disassembly-based bill-of-material database for 95 consumer electronic products. This provides valuable data about the materials and components constituting these devices, which are very rarely disclosed as information. However, very few IoT devices are included and substantial information is lacking to perform LCA for electronics components, e.g. die area values and integrated circuits (ICs) inventory. This should be considered for further work and especially for emerging categories such as IoT devices. A comprehensive study \cite{malmodin2018energy} focusing on the energy and carbon footprint of ICT over the period 2010-2015 identifies IoT as a remaining challenge for assessing ICT global footprint. They propose a macroscopic analysis in order to quantify the additional footprint due to IoT deployment in 2015. Their conclusion is that it only adds marginally to the overall carbon footprint of ICT, despite the large volumes of equipment and devices that could potentially be connected. However, details on modeling assumptions for IoT devices are not clearly presented. This will be further discussed in Section \ref{sec:macroIoT}. \\

More specific analyses are carried out in \cite{van2015life_smarttextiles, gurova2020sustainable} which point out the difficulty to quantify the environmental impacts of IoT in emerging applications such as smart textiles \cite{schischke2020-textiles}. In \cite{van2015life_smarttextiles}, the authors attempt to quantify the impacts of the electronic parts through the eco-cost indicator \cite{vogtlander2001virtual}. It can be seen in their supplementary material that ICs have not been considered nor magnets in the vibration motor. Moreover, Ecoinvent and Idemat databases have been used even though they are often pointed in the literature as not the best suited for LCA of electronics \cite{ercan2016, louis2020sources}. This is also the case in \cite{kokare2020comparative} for LCA of cardiac monitoring devices. Nevertheless, these are interesting attempts to better account for the additional impacts due to the presence of electronic components. A methodology to assess the direct environmental impacts of wireless sensors networks (WSN) is presented in \cite{bonvoisin2012}. It underlines the necessity to adopt a system-level perspective while performing environmental assessment of ICT. They conclude that energy consumption of WSN devices is of primary importance. However, even if edge devices seem to be responsible for the main part of the impacts, hardware is quickly described especially for electronics which is simply described by total mass. An integrated framework supporting the design of environmentally-sound ICT-based optimization services is proposed in a second paper by the same authors \cite{bonvoisin2014}. Carbon footprint and resource depletion results are given for each WSN but no more details are available which makes comparison and re-uptake difficult. Typical breakdowns of the carbon footprint for IoT device are given in \cite{bol2013green, bol2015can, bol2011application}, with respect to hardware components such as PCB, ICs and batteries. The results are mainly based on literature data and the authors highlight that application-awareness is critical in semiconductor LCA. Guidelines are proposed for the sustainable deployment of a large number of WSN. It requires minimizing (i) the embodied energy and carbon footprint of the WSN production, (ii) the ecotoxicity of the WSN e-waste, and (iii) the Internet traffic associated to the generated data. The challenge of eco-design in the IoT and the complexity of shared infrastructures is pointed out in \cite{quisbert2020lifecycle}. The authors argue that the life-cycle model of an IoT system should include the modeling of sensing, edge and cloud devices together with a representation of the hardware components. Although this work is of great interest because it aims at encompassing the full architecture of IoT infrastructure, no quantitative results are provided for the use case example. Nevertheless, this is the only work combining hardware details with life-cycle considerations for IoT, to the best of our knowledge. The lack of comprehensive study on the current and future footprint of IoT is also discussed in \cite{teehan2014_PhD, freitag2021climate, Gray2018}. It is pointed out that integrated analyses or meta-analyses are of greater utility than point estimates as new products continue to emerge on the market and span formerly distinct product categories \cite{teehan2014_PhD}. This motivates a shift towards broader frameworks considering multiple products simultaneously.


\section{Methodology} \label{sec:methodology}

This section describes the framework and the modeling assumptions used to generate the results. Since it is infeasible to consider every IoT design for every IoT application and since performing a single LCA for a specific IoT device would not be representative of the high diversity of designs and applications in this field, we introduce a framework that aims at streamlining the cradle-to-gate carbon footprint of several IoT devices. To tackle this complexity, we propose a bottom-up approach based on \textit{IoT hardware profiles}.

\subsection{IoT hardware profiles}

\begin{figure*}[h!]
    \centering
    \includegraphics[width=0.85\textwidth]{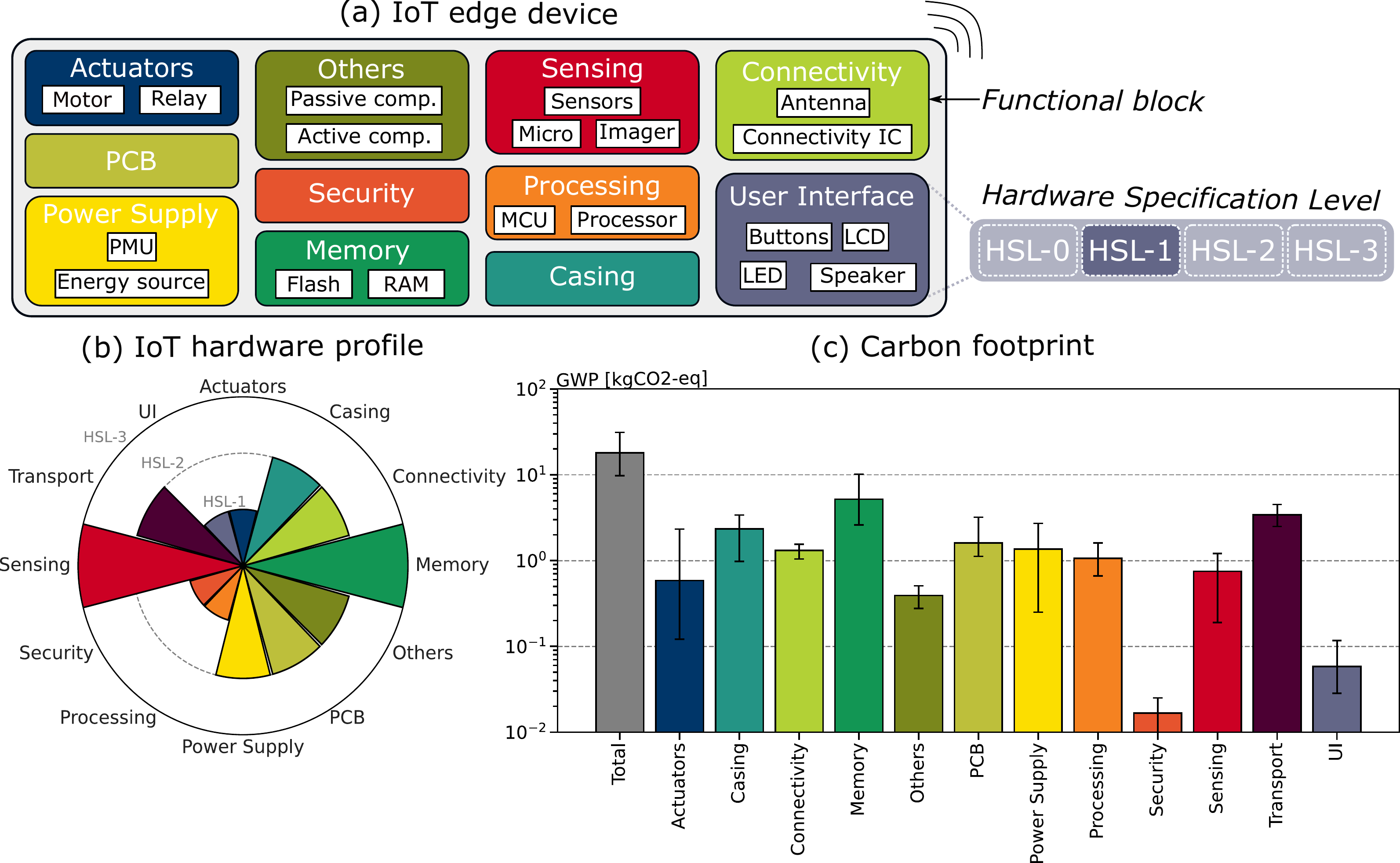}
    \caption{(a) The general architecture of an IoT edge device with the concept of \textit{functional blocks} and \textit{hardware specification level}, (b) an example of IoT hardware profile, (c) the resulting carbon footprint obtained by the framework for the hardware profile in (b).}
    \label{fig:IoT}
\end{figure*}

In our framework, an IoT hardware profile is the description of an IoT edge device in terms of \textit{functional blocks}, leveraging the similarities between IoT edge devices while maintaining some flexibility through the \textit{hardware specification level}. The \textit{functional block} gathers components that perform a common function as represented by the general architecture of an IoT edge device presented in Fig. \ref{fig:IoT}(a). It has been inspired by ETSI \cite{ETSI} and ITU \cite{ITU} standards for the LCA of ICT equipments even though we do not claim full compliance due to the parametric nature of our study. The \textit{hardware specification level} (HSL) gives details about the hardware available within a \textit{functional block}. The classification is structured along these two axes, as detailed in Table \ref{table_IoT_details}. This can then be used to define hardware profiles as shown in Fig. \ref{fig:IoT}(b). The different levels proposed in Table \ref{table_IoT_details} are based on the analysis of several available IoT edge devices. While this list is non-exhaustive, it already covers a wide variety of current IoT edge devices. Other classifications of IoT exist \cite{gubbi2013internet} but they are usually more focused on IoT applications rather than on the hardware itself, as proposed in \cite{sparks2017route_ARM, samie2016iot}. \\

In order to provide the aforementioned flexibility, LCA has been carried out for each \textit{hardware specification level}. The total shown in Fig. \ref{fig:IoT}(c) is given by the sum of the carbon footprint contributions for a given hardware profile : the result is therefore uniquely defined by the hardware profile of the IoT edge device.

\clearpage
\onecolumn

\begin{landscape}
\begin{table*}
\setlength\tabcolsep{4pt}
\caption{Detailed life-cycle inventory (LCI) for IoT hardware profiles. For each \textit{Hardware Specification Level}, lower/typical/upper parameters considered are given.}
\vspace{0.5cm}
\label{table_IoT_details}
\renewcommand{\arraystretch}{1.35}
\centering
\resizebox{22cm}{!}{

\begin{tabular}{l l l l l}

\toprule[2pt]

 & \multicolumn{4}{c}{\LARGE{\textbf{Hardware Specification Level (HSL)}}}\\
 \cmidrule{2-5}
 \LARGE{\textbf{Functional}} & \Large{HSL-0} & \Large{HSL-1} & \Large{HSL-2} & \Large{HSL-3} \\
\LARGE{\textbf{Block}} & \multicolumn{4}{l}{}\\
\midrule[2pt]

\textbf{Actuators} & No actuator & Vibration motor (1g) 1/2/4 & Relay (SSR) 1/2/4  & DC motor (50g) 1/4/6 \\
& & & & Motor driver $^\ddagger$ 1/2/5 $mm^2$ \\
& & & & Motor driver transistor 1/4/6 \\

\midrule[0.2pt]

\textbf{Casing} & No casing & ABS plastic granulate 10/50/100 g & ABS plastic granulate 200/400/500 g & ABS plastic granulate 700/800/900 g \\
& & Aluminium 1/10/30 g & Aluminium 20/80/150 g & Aluminium 70/160/300 g \\
& & Steel 1/10/30 g & Steel 20/80/150 g & Steel 70/160/300 g \\

\midrule[0.2pt]

\textbf{Connectivity} & Embedded in \textit{Processing} (share of the die area) & Connectivity IC $^*$  5/10/20 $mm^2$ & Connectivity IC $^\blacktriangle$  20/30/45 $mm^2$ & Connectivity IC $^\blacktriangle$  45/50/60 $mm^2$ \\

& Printed antenna (embedded in \textit{PCB})  & Printed antenna (embedded in \textit{PCB})  & External whip-like antenna 10/15/30 g  & External whip-like antenna 10/15/30 g \\

\midrule[0.2pt]

\textbf{Memory} & Embedded in \textit{Processing}, Flash + RAM ($\simeq$ kB) & DRAM $^\diamond$  (32/128/512 MB) 2/7.9/31.5 $mm^2$ & DRAM $^\diamond$ (0.5/1/2 GB) 31.5/61.5/123.1 $mm^2$ & DRAM $^\diamond$ (0.5/1/2 GB) 31.5/61.5/123.1 $mm^2$  \\
& & Flash $^\dagger$ (32/128/512 MB) 0.2/0.8/3.2 $mm^2$ & Flash $^\dagger$ (1/4/8 GB) 6.3/25/50 $mm^2$ & Flash $^\dagger$ (8/16/32 GB) 50/100/200 $mm^2$  \\

\midrule[0.2pt]

\textbf{Others} & Capacitors and resistors 5/10/15 & Capacitors and resistors 15/20/25 & Capacitors and resistors 40/50/60 & Capacitors and resistors 75/85/100 \\
& Diodes 2/2/2, transistors 1/2/3 & Diodes 2/4/6, transistors 2/4/6 & Diodes 2/4/6, transistors 4/7/9 & Diodes 2/6/10, transistors 7/10/15 \\
& Tantalum capacitors 0/0/2, crystals 0/1/1 & Tantalum capacitors 0/0/3, crystals 1/1/2 & Tantalum capacitors 0/0/4, crystals 1/2/4 & Tantalum capacitors 0/2/4, crystals 1/2/4  \\
& & Steel metal shield 0.5/1/2 g, cables 1/2/5 cm & Steel metal shield 0.5/1/2 g , cables 1/2/5 cm & Steel metal shield 0.5/1/2 g, cables 1/2/5 cm \\

\midrule[0.2pt]

\textbf{PCB} & FR4 (4 layers) 8/10/15 $cm^2$ & FR4 (4 layers) 15/35/50 $cm^2 $ & FR4 (8 layers) 35/50/100 $cm^2$ & FR4 (8 layers) 80/120/150 $cm^2$  \\  
& Solder Paste (SAC305) 4/8/13 mg  & Solder Paste (SAC305) 28/53/98 mg & Solder Paste (SAC305) 99/155/249 mg & Solder Paste (SAC305) 178/265/454 mg \\
 
\midrule[0.2pt]
 
\textbf{Power Supply} & Mains powered & 1 Coin cell Li-Po/2 AAA alkaline/2 AA alkaline & Li-ion battery 10/50/100 g &  Li-ion battery 10/50/100 g\\
& Power transistor 2/3/4 & & Power transistor 0/1/2 & Power transistor 0/1/2  \\
& Diodes power 0/1/2, radial capacitor 2/3/4 & & Diodes power 0/1/2, radial capacitor 0/1/2 & Diodes power 0/1/2, radial capacitor 0/1/2 \\
& Miniature coil 2/3/4, ring core coil 0/1/1 & & Miniature coil 0/1/2 & Miniature coil 0/1/2   \\
& Power cable 0.5/1/1.5 m & & & External IC $^\ddagger$ 5/15/25 $mm^2$ \\
& CEE 7/4 Schuko plug 0/1/1 & & &  \\

\midrule[0.2pt]

\textbf{Processing} & MCU $^*$ 5/10/17 $mm^2$ & Application processor $^\vartriangle$ 20/30/45 $mm^2$ & Application processor $^\vartriangle$ 50/60/75 $mm^2$ & Application processor $^\vartriangle$ 75/100/125 $mm^2$ \\
& & Auxiliary MCU $^*$ 5/10/17 $mm^2$ & Auxiliary MCU $^*$ 5/10/17 $mm^2$ & Auxiliary MCU $^*$ 5/10/17 $mm^2$ \\

\midrule[0.2pt]

\textbf{Security} &  Embedded in \textit{Processing} or non-existent & External IC $^\ddagger$ 1/2/3 $mm^2$ & N/A & N/A \\

\midrule[0.2pt]

\textbf{Sensing} & No sensor & Electret microphone 0.05/0.1/0.2 g & Single-multiple sensors $^\circ$ 0/3/5 $mm^2$ & Single-multiple sensors $^\circ$ 0/3/5 $mm^2$ \\
& & & & Single CMOS imager $^\ddagger$ (1/4” to 2/3”) 8/30/58 $mm^2$ \\

\midrule[0.2pt]

\textbf{Transport} & No transport & Transport from China to Europe & Transport from China to Europe & Transport from China to Europe \\
    & & Truck distance : 100/300/600 km & Truck distance : 100/300/600 km & Truck distance : 600/900/1200 km \\
    & & Plane distance : 6100/6775/7400 km & Plane distance : 6100/6775/7400 km & Plane distance : 6100/6775/7400 km \\
    & & Total weight = 50/100/300 g & Total weight = 300/650/900 g & Total weight = 900/1500/2000 g \\

\midrule[0.2pt]

\textbf{User Interface} & No user interface & Switch-button 0/1/2 & Switch-button 0/2/3 & Switch-button 2/3/4 \\
& &  LED 1/2/4 & LED 2/4/6, LED driver $^\ddagger$ 0/1/2 $mm^2$ & LED 3/5/8, LED driver $^\ddagger$ 0/1/2 $mm^2$ \\
& & &  Speaker 2/10/40 g, audio driver $^\ddagger$ 1/2/5 $mm^2$  & 1 LCD screen 5/25/100 $cm^2$, driver $^\ddagger$ 0/1/2 $mm^2$. \\

\bottomrule[2pt]

\end{tabular}
}
\vspace{0.1cm}
\flushleft{\qquad \small{
$^\circ$ CMOS $0.25\mu m$
$^\ddagger$ CMOS $0.13\mu m$
$^*$ CMOS 90nm
$^\blacktriangle$ CMOS 22nm
$^\vartriangle$ CMOS 14 nm
$^\dagger$ Flash 45nm
$^\diamond$ DRAM 57nm \\
}}
\end{table*}

\end{landscape}



\subsection{Modeling assumptions} \label{sec:modeling}

\begin{figure*}[ht!]
    \centering
    \includegraphics[width=1\textwidth]{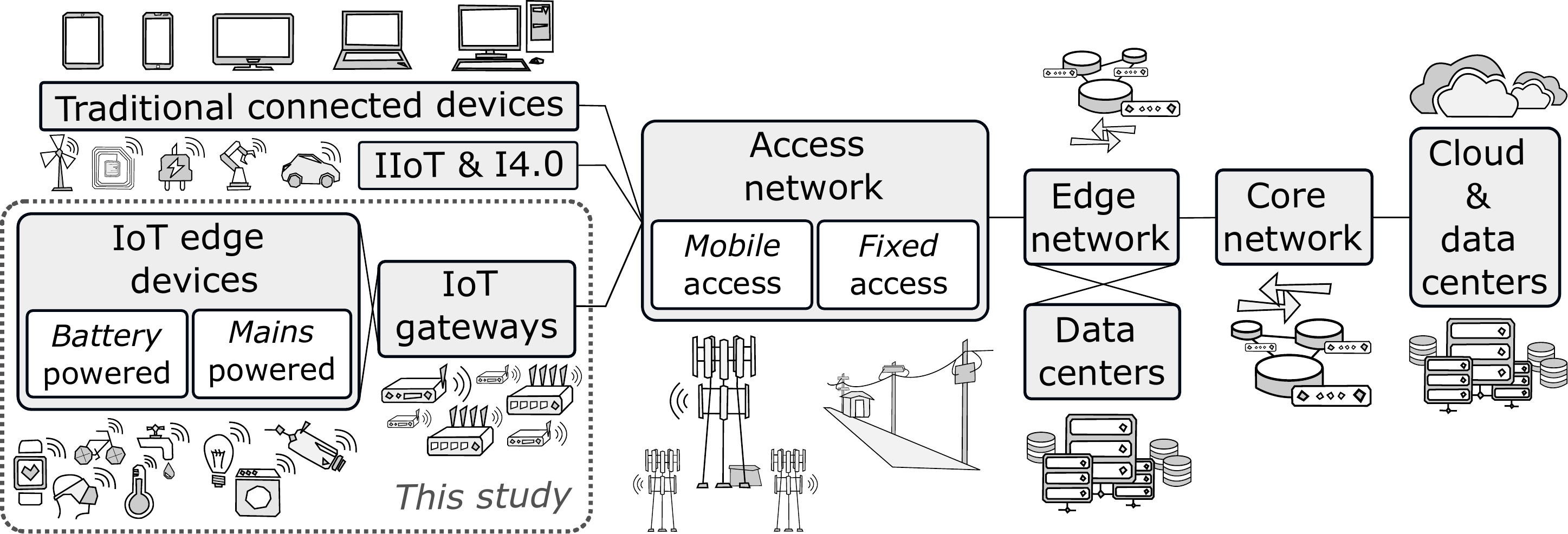}
    \caption{Schematic IoT network representation, adapted from \cite{EDNA2016, Gray2018, samie2016iot}. This study focuses on the IoT edge devices and gateways.}
    \label{fig:IoT_edge}
\end{figure*}

This study is built up on the LCA methodology described by ISO14040 standards series. \textbf{Goal and scope:} The goal of this study is to have a better understanding of the carbon footprint of the production of a wide range of IoT edge devices. Several hardware profiles are considered to illustrate the diversity of IoT edge devices and point out the variability existing between different applications and designs. It also allows a streamlined yet specific assessment for cradle-to-gate life cycle impacts of a given IoT edge device. \textbf{Functional unit:} The functional unit is \textit{the production and transport to the use location of a single IoT edge device defined by its hardware profile}. \textbf{System boundaries:} This study considers a cradle-to-gate analysis including life cycles stages and processes from raw material extraction and production. Transport to the use location is taken into account. Only IoT edge devices are covered in this study as depicted in Fig. \ref{fig:IoT_edge}. The use phase, the associated usage of networks and infrastructures and the end-of-life are not covered by this framework. Some cut-offs have been applied: connectors, chargers, final packaging materials and final assembly are not taken into account in this framework. \textbf{Impact categories:} This study focuses on the carbon footprint and considers only the global warming potential (GWP) midpoint indicator with the ReCiPe 2016 v1.1 (H) method. \\

Standards motivate the use of specific data with respect to time, geography and technology \cite{ETSI}. However, primary data is extremely hard to access for ICT devices and electronic components in general. Moreover, as the production of electronics components is known to be a very intensive process in terms of resources and energy \cite{boyd2011life, louis2020sources}, great care should be taken when mapping hardware components to impact factors or entries in databases. In this work, we use the GaBi Electronics XI database \cite{GaBiElecDB} on top of the Professional GaBi database, for its high level of details on electronic components and its extensive documentation. Other sources are used such as Ecoinvent v3.5, Base Carbone and literature (mostly for comparison).  \par

For all ICs, the estimation of silicon die area is based on an analysis of datasheets for IoT edge devices components. Scaling of the database entries are based on the die-to-package ratio given in the GaBi documentation and the targeted die size indicated in Table \ref{table_IoT_details}. All ICs in this study are modeled using the GaBi Electronic XI database which is based on their parametric IC model. Production yields \cite{boyd2011life} are considered and the model covers main production aspects: silicon wafer production, front-end (die manufacturing) and back-end (packaging) processing, production of gases, chemicals and materials, perfluocarbon (PFC) emissions, and facility impacts. \\

The following sub-sections details the modeling assumptions for each \textit{functional block}. Additional details can be found in the supplementary material. 

\subsubsection{Actuators}

\textit{Block function:} components that physically act on the environment.\par

\textit{Hardware specifications:} this framework models three type of actuators. A relay, a vibration motor and a small DC motor. As none of them are available in the GaBi database, custom models were developed. The Solid State Relay (SSR) is modeled using an LED, a power transistor and a $1 mm^2$ CMOS control circuit. For both motors, we consider NdFeB magnet by scaling another entry of GaBi data base i.e. a 2g speaker, see Section \ref{sec:UI-details} for more details. In addition, we consider steel, copper, a small 2-layer PCB and a small driver for the vibration motor. Regarding the DC motor, we include copper for wiring and coils, steel and thermoplastic for the bearings and structural components. A small IC is also included as motor control and driver with power transistors. \par

\subsubsection{Casing}

\textit{Block function:} structural components of the IoT edge device whose primary function is to protect the electronic components. \par

\textit{Hardware specifications:} this framework models the casing as a mix of Acrylonitrile-Butadiene-Styrene Granulate (ABS) together with a process of plastic injection moulding supplied by the Chinese energy mix. Losses in the injection moulding process are considered up to 3\%. Aluminium contribution is modeled by considering aluminium profiles produced through an extrusion process. Steel ingots model the presence of steel but no extrusion or casting process is considered. A ratio of 10-20-30\% between aluminium-steel and plastic is supposed based on a rough analysis of the data available in \cite{babbitt2020disassembly_nature} even though this is far more variable in practice. The impact of material selection for the casing is discussed in \cite{louis2020sources}. \par

\subsubsection{Connectivity}

\textit{Block function:} components which are involved in data transmission. This is a critical and necessary feature for IoT edge devices. Many protocols and technologies exist for IoT communication such as ZigBee, NB-IoT, LoRa, Bluetooth and WiFi \cite{samie2016iot, lethaby2017_texas_IoT}. Each one comes with different trade-off in terms of range, data rate, and power consumption. \par

\textit{Hardware specifications:} this framework models connectivity hardware based on an equivalent silicon die area through CMOS 90 nm or 22 nm technology node. PCB printed antennas are considered to be part of the PCB \textit{functional block}. For external antenna, we use a basic custom model based on a mix (20-40-40\%) of copper, steel and ABS to model a 15g whip-like antenna. Regarding connectivity, some protocols or network architectures need a gateway for proper operation. This can be taken into account by considering the gateway as an IoT edge device and applying this framework to the gateway itself, thereby allocating its own impacts to the IoT edge devices that use it. Note that in Table \ref{table_IoT_details}, the lowest hardware specification level (HSL-0) has connectivity capabilities directly embedded in the Processing \textit{functional block}, such as in a wireless microcontroller with embedded radio transceiver. \par

\subsubsection{Memory}

\textit{Block function:} components involved in the data storage on the IoT edge device.  \par

\textit{Hardware specifications:} this relates only to stand-alone external memory ICs. A distinction between volatile (e.g DRAM) and non-volatile memory (e.g. Flash) is done. This framework models memory hardware based on an equivalent silicon die area and CMOS 45/57 nm technology node, for Flash and DRAM respectively. Memory density is considered when actual die size is not available: 0.13 Gb/$mm^2$ is used for (S/D)RAM and 1.28 Gb/$mm^2$ for Flash \cite{Schischke2021-EUreport}. Additional details are given in the supplementary material. \par

\subsubsection{Printed circuit board (PCB)}

\textit{Block function:} components responsible for the electro-mechanical support and electrical connections between electronic components of the IoT edge device. \par

\textit{Hardware specifications:} this framework models PCB hardware based on area and number of layers, as suggested in ETSI and ITU standards. The solder paste used to fasten components to the PCB is also taken into account by considering SnAg3Cu0.5 (SAC) solder paste. The quantity is defined according to the total area of ICs, considering a thickness of $0.1 mm$ and a mass density of $7.38 g/cm^3$ \cite{solder_paste}. As mentioned in GaBi documentation for rigid PCB with HASL finish, a subtractive method is considered with relevant process steps  through a parametrized model e.g. core preparation, layering, outer layer preparation, finishing and overheads. Assembly for through-hole technology (THT) and surface-mount device (SMD) components are also taken into account in GaBi (600/h throughput). \par

\subsubsection{Power supply}

\textit{Block function:} components involved in the energy source and energy management of the IoT edge device, which could be battery-powered or main-powered. \par

\textit{Hardware specifications:} this framework models power supply hardware through coin cell, AA/AAA alkaline batteries or Li-ion batteries. Coin cell and AA/AAA alkaline batteries are taken directly from GaBi. However, GaBi does not contain Li-ion components. Results from another study \cite{ellingsen2014_LCA_Lion} have been used as a proxy to model carbon footprint of Li-ion battery production. Their results were normalized by weight and capacity which is straightforward to scale. Carbon footprint is similar to \cite{ercan2016} which used primary data but did not detail the life cycle inventory for the battery. In fact, 1.4 kgCO2-eq for a 48g Li-ion battery gives 29.17 kgCO2-eq/kg, close to the 25 kgCO2-eq/kg from \cite{ellingsen2014_LCA_Lion}. In \cite{Proske_FF3}, an emission factor of 29.62 kgCO2-eq/kg is used. Ecoinvent data for Li-ion battery ranges from 5.7 to 7.7 kgCO2-eq/kg but it is from 2011 and consequently outdated. The framework also takes into account a power management unit (PMU) for main-powered edge devices by considering basic AC/DC and DC/DC converters through coils, capacitors, power transistors and diodes. A power cord and a CEE 7/4 Schuko plug model are also considered. Some IoT edge devices use a dedicated PMU IC which is captured in HSL-3. \par

\subsubsection{Processing}

\textit{Block function:} components involved in the data processing and control tasks on the IoT edge device. \par

\textit{Hardware specifications:} this framework models processing based on an equivalent silicon die area through CMOS 90nm technology node for microcontrollers (MCU) and CMOS 14nm for application processors. Common IoT edge devices often use Cortex-M MCUs and Cortex-A applications processors \cite{adegbija2017microprocessor, samie2016iot}. Moreover, small embedded memory in the main MCU and processor is considered.\par

\subsubsection{Security}

\textit{Block function:} components related to cryptography and protection of sensitive data. \par

\textit{Hardware specifications:} this covers only external stand-alone security ICs. In practice, most IoT edge devices embed this function directly in the main MCU and processor e.g. embedded AES encryption. This framework models security hardware based on an equivalent silicon die area and CMOS 90nm technology node. \par

\subsubsection{Sensing}

\textit{Block function:} components involved in measuring physical quantities of the environment. Some common sensors are temperature, pressure and audio, proximity, inertial measurement unit (IMU), passive infrared sensors (PIR), GPS, image and video, among others. \par

\textit{Hardware specifications:} this framework models sensors based on an equivalent silicon die area through CMOS $0.25\mu m$ technology node. For audio sensor, an electret condenser microphone from GaBi is considered. For camera, only the CMOS sensor is modeled by an equivalent silicon die area in $0.13\mu m$ as it represents the main part of the impacts \cite{andrae2017precision}: sensor area range from $1/4"$ to $2/3"$. \par

\subsubsection{Transport}

\textit{Block function:} related to the shipping of the assembled IoT edge device from the factory to the consumer.  \par

\textit{Hardware specifications:} transport is modeled based on chargeable weight as motivated in ETSI-ITU standards. It is assumed that the IoT device is sent from China to Europe by plane and then transported by truck to its destination in Europe. \par

\subsubsection{User interfaces (UI)} \label{sec:UI-details}

\textit{Block function:} components allowing interactions between external users and the IoT edge device. \par

\textit{Hardware specifications:} this framework models user interfaces through switch buttons, LEDs, a speaker and an LCD display. For the display, panel size is ranging from $5 cm^2$ to $100 cm^2$. For the sake of comparison, a 2020 smartphone screen is about $80-90 cm^2$. As GaBi databases does not contain any display modules, modeling is done using primary data for touch screens \cite{AUO}, similarly to \cite{Proske_FF3}.  Push buttons and LEDs are taken from GaBi. A small CMOS driver is considered for the LEDs. The original 2g speaker with neodymium selected in GaBi is proportionally scaled up on a weight basis to cover heavier speakers: the resulting emission factor per kg seems coherent with other studies focusing on NdFeB magnets \cite{marx2018comparative}, respectively 57.3 kgCO2-eq/kg compared to 33.5 kgCO2-eq/kg in the Base Carbone and 41.4-89.2 kgCO2-eq/kg in \cite{marx2018comparative}. A small audio driver is also considered. \par

\subsubsection{Others}

\textit{Block function:} the rest of the components which do not fit another \textit{functional block}. \par

\textit{Hardware specifications:} this includes capacitors, resistors, coils, crystal, additional transistors, electromagnetic shields (EMS) for ICs and cables. \par


\section{Results} \label{sec:results}  \label{sec:use-cases}

This section presents the carbon footprints obtained by applying the hardware framework to four use cases. Results are then interpreted and compared with existing benchmarks when applicable. Finally, a sensitivity analysis is carried out for the framework. 

\subsection{Carbon footprint of IoT edge devices : use cases}

Figure \ref{fig:results_all} illustrates the results for all use cases analyzed in this study. For each \textit{functional block}, the uncertainty range is given according to the lower-typical-upper parameter defined in Table \ref{table_IoT_details}.

\subsubsection{Occupancy sensor}

A complete teardown has been carried out on the occupancy sensor from Philips HUE. A small Flash memory (4 Mb) is present on the edge device but not captured by the hardware profile. This is a typical example of a light hardware profile. As this IoT edge device needs a gateway to function properly, an additional hardware profile could be generated for the gateway by using our framework.

\subsubsection{Drone}

Teardown reports available on-line have been used for the DJI MAVIC mini which is a light-weight drone of less than 250g. The remote controller is not considered although a dedicated hardware profile could also be generated for it. The smartphone used to acquire real-time data from the drone has not been taken into account either. The propellers usually made out of glass-fiber-reinforced composite or carbon-fiber-reinforced material are not captured by the hardware profile. In addition, some ICs were impossible to identify and therefore discarded, as detailed in the supplementary material.

\subsubsection{Connected home assistant}

A complete teardown has been carried out on the Google Home mini which is the light-weight version of the Google Home assistant. The silicon die area for ICs has been obtained through die inspection. We used a chemical destructive method which attacks selectively the epoxy package and not the silicon die. More details are available in supplementary material. The carbon footprint of the loudspeaker should be investigated more deeply. Indeed, the loudspeaker weights about 80g which is not properly captured by its hardware profile. External devices used to connect to the home assistant have not been taken into account. 

\subsubsection{Smart watch}

Teardown reports available on-line have been used for Apple smart watches and the Garmin Fenix 5 smart watch. Several components do not have a datasheet or are impossible to identify which prevents precise inventory. This is the case for some ICs or TAPTIC engines in Apple's products for instance. External devices used to connect to the smart watch have not been taken into account. 

\subsection{General interpretation} \label{sec:intepretation}

\begin{figure*}
    \centering
    \includegraphics[width=1\textwidth]{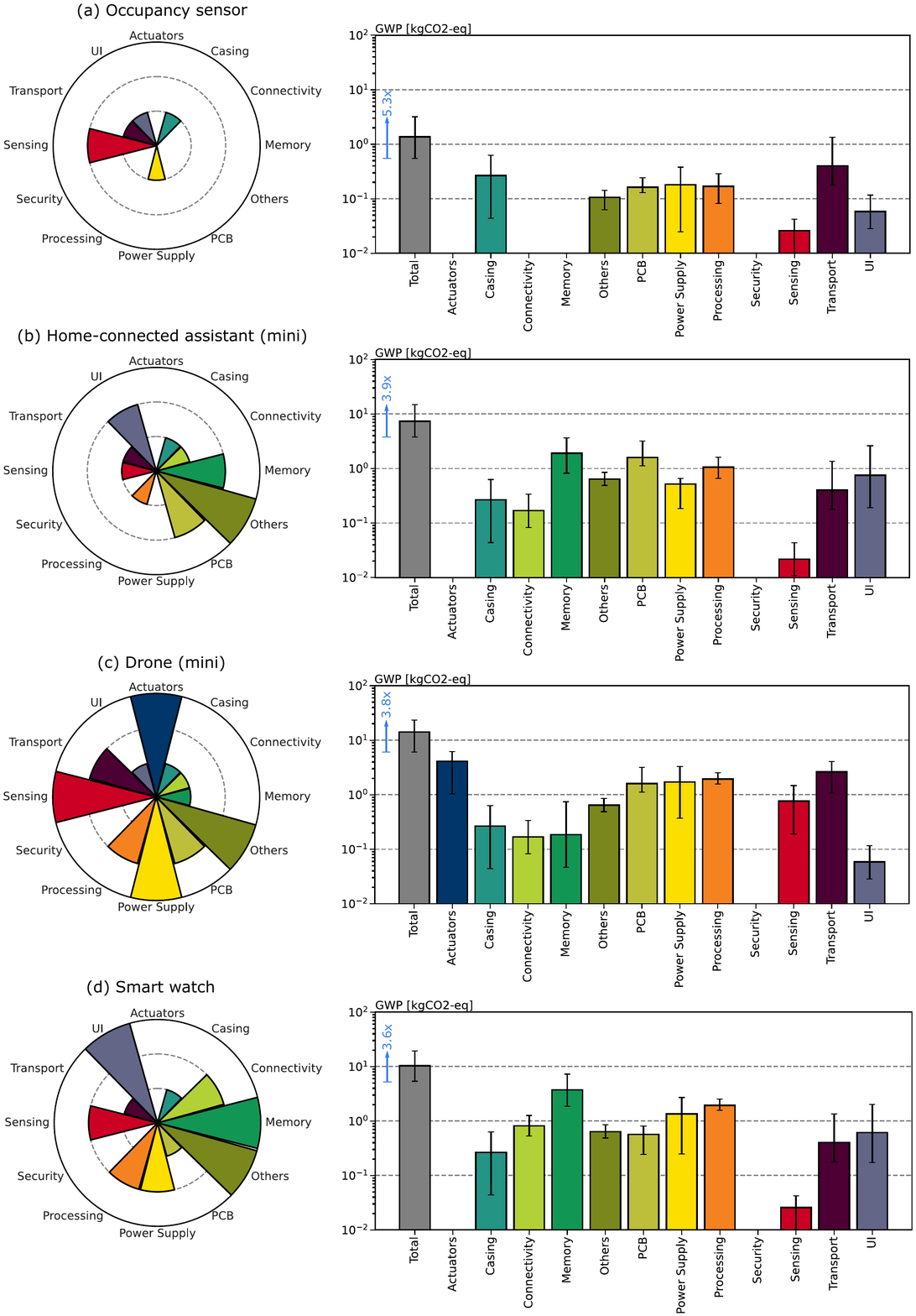}

    \caption{Carbon footprint results based on IoT hardware profiles for the use cases considered in this study: (a) a Philips HUE occupancy sensor, (b) a Google Home mini home connected assistant, (c) a DJI MAVIC mini light-weight drone and (d) a smart watch from Apple and Garmin.}
    \label{fig:results_all}
\end{figure*}

For simple IoT edge devices with limited functionality such as the occupancy sensor, the resulting cradle-to-gate carbon footprint is about 1.4 kgC02-eq (0.6 to 3.2 kgCO2-eq) which is a fairly low footprint. For more complex IoT nodes, the cradle-to-gate carbon footprint increases : 3.8/7.3/14.9 kgCO2-eq for the home-connected assistant, 6.1/14.1/23.4 kgCO2-eq for the drone and 5.4/10.4/19.5 kgCO2-eq for the smart watch. As depicted by the total bars in Fig. \ref{fig:results_all}, the ratio between the upper and lower values for each result can be up to a factor 5$\times$. This can be further reduced for each use case by a more in-depth analysis using Table \ref{table_IoT_details} and the bar plots of each \textit{functional block}.  \\

For simple devices with light hardware profiles, the main part of environmental impacts is caused by the non-electronic parts such as the casing and transport as well as production of the battery and the PCB rather than by ICs production. For heavier hardware profiles however, ICs production is responsible for a significant part of the carbon footprint, together with the PCB. Actuators and casing can in some cases be responsible for an important share of the carbon footprint which motivates not to consider only electronic components in the LCA of IoT. The connectivity \textit{functional block}  has a low responsibility in terms of embodied impacts even if it is a critical feature for IoT devices. The low contribution of displays is mainly due to the small screen sizes considered. The reasoning is similar for camera sensors.
\clearpage 

\clearpage 

\subsection{Comparison with existing studies} \label{sec:comparison}

\begin{figure*}[h!]
    \centering
    \includegraphics[width=1\textwidth]{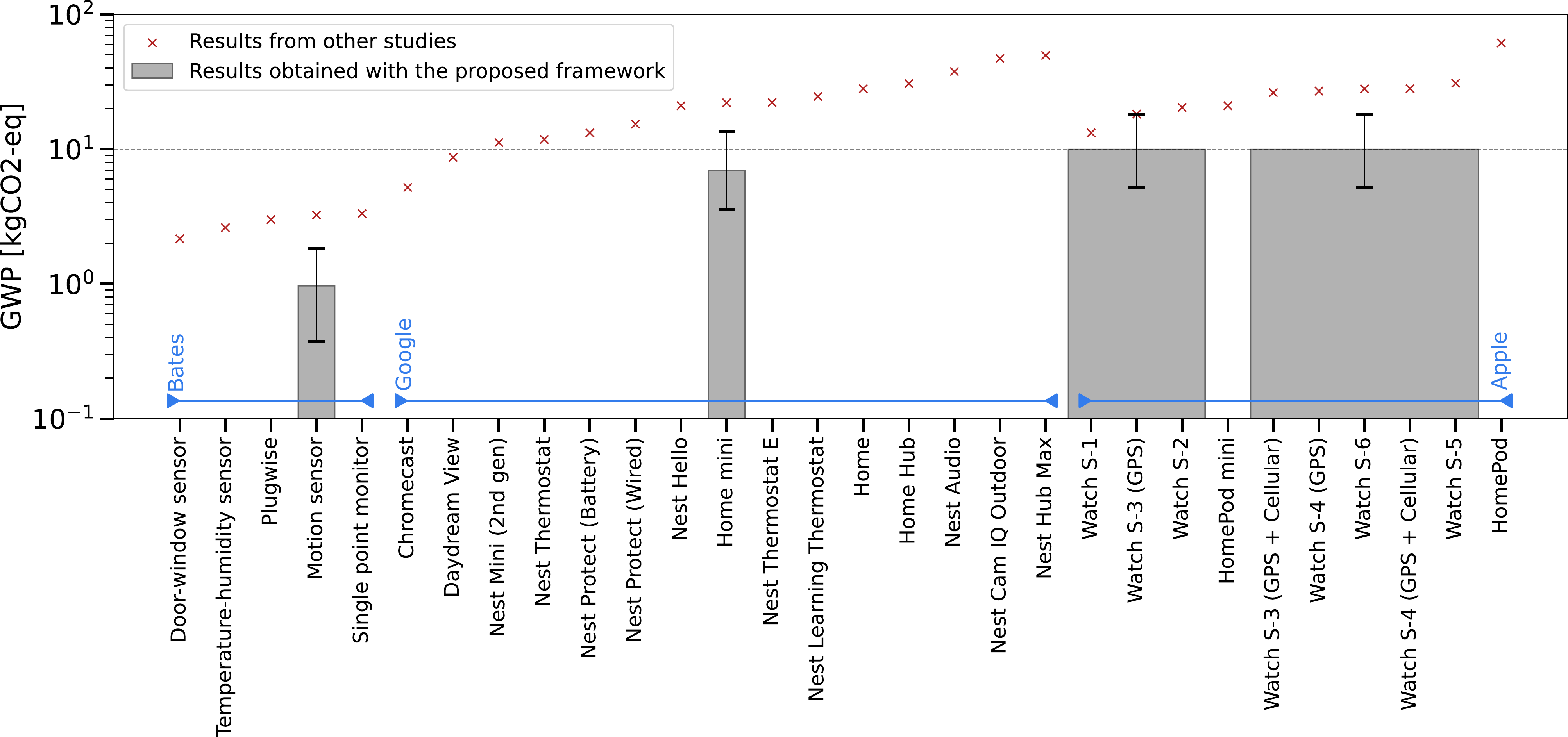}
    \vspace{-0.6cm}
    \caption{Comparison between the carbon footprints obtained for the use cases in this work and independent reports or studies when available i.e. occupancy sensor \cite{bates2013exploring}, home connected assistant \cite{GoogleSust}, and smart watch \cite{Apple_PER}. Only production is shown for clarity. No benchmark found for the drone. }
    \label{fig:IoT_available_reports}
\end{figure*}

Very few LCA results of IoT devices are available for comparison. The carbon footprints of several sensors (e.g. motion and temperature-humidity sensors) are given in \cite{bates2013exploring} but the study is from 2013 (likely outdated) and very few details are available. Google \cite{GoogleSust} and Apple \cite{Apple_PER} have published the carbon footprint of some of their products including connected assistants and smart watches. Figure \ref{fig:IoT_available_reports} shows the comparison between the results of our framework and the aforementioned studies. It provides very similar trend and accounts for the variation between devices but the absolute results given in environmental reports are usually higher than those predicted by this framework, up to a factor $3\times$ for production and $2\times$ for transport. A similar comparison with Apple's LCA results was done for traditional connected devices (e.g. laptops, netbook, tablets and music player) \cite{teehan2013comparing} and an undershoot factor ranging from $1.3 \times$ up to $2.7 \times$ was identified. The authors pointed out that bottom-up approaches suffer from truncation errors that can explain the undershoot \cite{teehan2014_PhD}, at least partially. Indeed, some internal specific processes are obviously unknown and consequently omitted here. The use of a different impact database by Apple could also partially explain the undershoot as the values for the IC carbon footprint in the GaBi databases are lower than the literature \cite{ercan2016, louis2020sources, teehan2014_PhD}. However, the lack of details in their reports make the comparison difficult, as also mentioned in \cite{teehan2014_PhD}. The more complex modeling of casing materials with an important share of metal and the presence of magnets in actuators seem to have a significant impact on carbon footprint. They were also identified as possible reasons to explain the undershoot. Regarding transportation, it is likely that more intermediate steps and travel methods occur in the logistics, thus impacting the distances and the associated footprint. If transportation packaging materials are considered, the total weight increases which also increases the carbon footprint.   

\subsection{Sensitivity analysis}

\begin{figure*}[h!]
    \centering
    \includegraphics[width=1\textwidth]{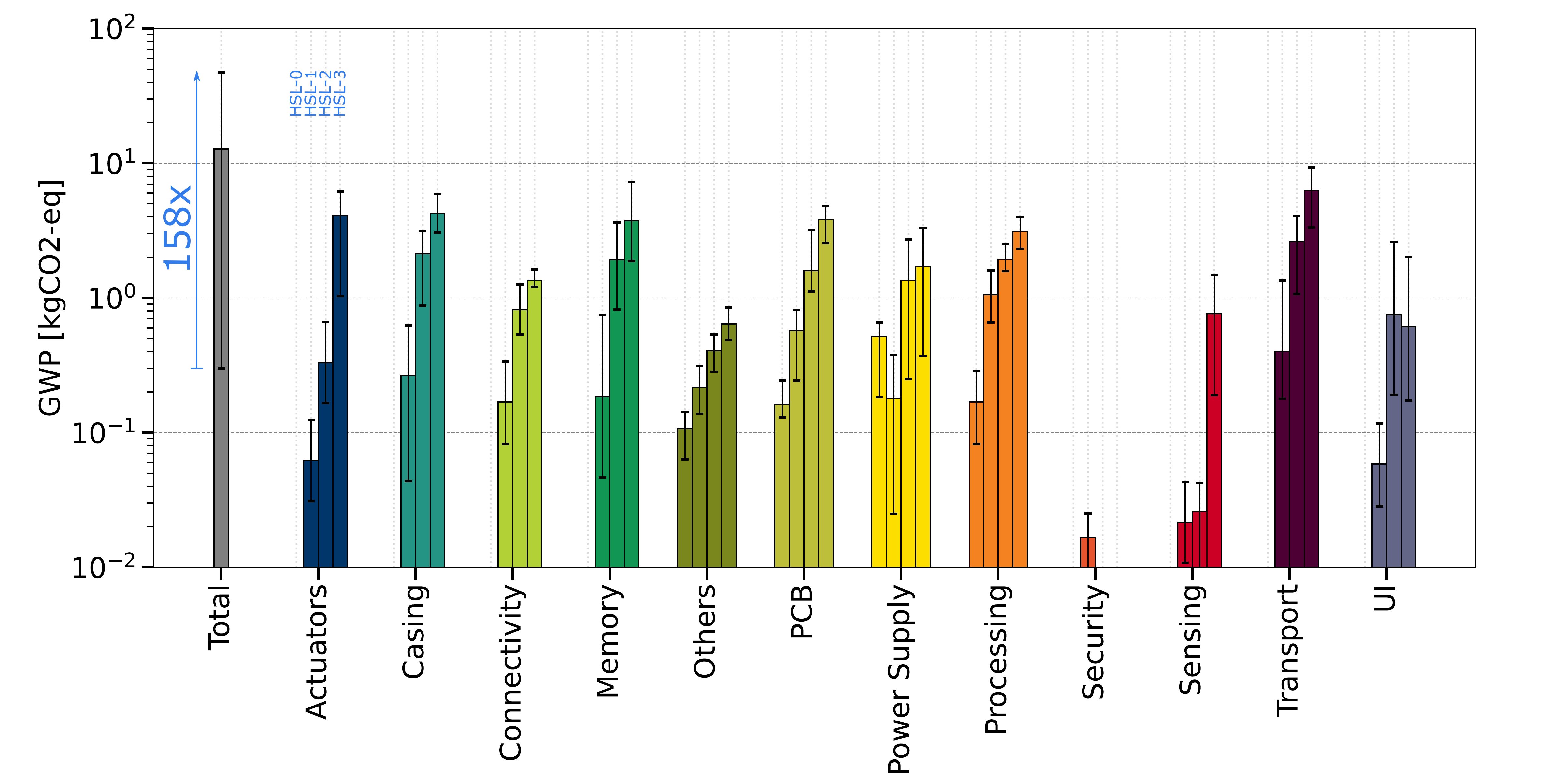}

    \caption{Sensitivity analysis across \textit{functional blocks} and \textit{hardware specification levels}. This shows the need to account for the heterogeneity in IoT edge devices as the production carbon footprint of simple and complex devices can be spread by a factor of $158 \times$, as modeled by our framework.}
    \label{fig:Sensitivity_analysis}
\end{figure*}

First, the stability of the results must be analyzed with respect to the modeling of the IoT edge devices. As the framework can model several combinations of \textit{functional blocks}, it is important to understand where the strongest impacts are. For each \textit{hardware specification level}, lower-typical-upper bounds have been considered as presented in Table \ref{table_IoT_details}. This can be seen as a sensitivity analysis based on parameter variation, which avoids results to be point estimates. We did not perform Monte Carlo analysis as it makes less sense for such a parametric framework than for a single specific device. Uncertainty is always clearly displayed in all results and the sensitivity of the results with respect to each \textit{functional block} and \textit{hardware specification level} has been analyzed as shown in Fig. \ref{fig:Sensitivity_analysis}. It shows that the ratio between the maximum and minimum carbon footprints obtained in the hardware framework can reach a factor of $158 \times$. This clearly underlines the wide range of carbon footprints for IoT edge devices with different hardware profiles. To put this in perspective, the ratio between maximum and minimum carbon footprints of smartphones lies usually below a factor $10\times$ \cite{louis2020sources}. As presented in Section \ref{sec:intepretation}, the results provided by our framework for a given hardware profile can still vary up to a factor 5$\times$. Although it is clear that it cannot outperform the precision of a dedicated LCA that would fully focus on one single IoT edge device, our framework provides streamlined carbon footprints for multiple products while maintaining a high resolution on electronic components.  \\

Then, it is important to discuss what could be the impact of internalizing limitations in the framework. Bulky casings in aluminium or steel and heavy actuators or speakers with magnets are not well covered in this framework and could increase the final footprint. This can be mitigated by re-evaluating the carbon footprint of the related \textit{functional block} with a ratio calculated on a weight basis. Another source of variation is the scaling of silicon die area based on the outer dimension of an IC chip package. The wide variety of packages and the variability in die-to-package ratio can lead to significant errors \cite{teehan2014_PhD, louis2020sources}. Indeed, using on-line teardowns to make the LCA of electronic products is more difficult and less precise because assumptions have to be made rather than using measurement data. Specifically, necessary information is not available in the datasheet of components or in existing on-line public teardowns \cite{louis2020sources, teehan2014_PhD}. This strongly limits the possibility to carry out LCA for IoT edge devices. Regarding the results sensitivity with respect to the cut-offs that have been introduced, variation due to final assembly is expected to be up to +10\% \cite{louis2020sources}. Connectors would have a limited impact with respect to GWP because gold usage contributes more to other impact categories. \cite{Proske_FF3, ercan2016}. Chargers could be modeled using the power supply \textit{functional block}, yet this would probably be an underestimation. \\

Finally, it is legitimate to question the sensitivity of the results with respect to the database that was used. Indeed, the discrepancies between the existing databases could impact the results. In this study, we made an extensive use of the GaBi Electronics database for its advanced modeling and level of details for electronic components. Regarding ICs, variation in the emission factor per die area is discussed in \cite{louis2020sources} for several academic sources. A deep analysis of the emission factors for different technology nodes comparing commercial databases, public data from industry-fabs and literature would be of great interest. Indeed, comparison between studies is hard and the lack of transparency prevents efficient re-uptake of the results \cite{louis2020sources}. In addition, some critical parts are not modeled in GaBi databases, such as WiFi modules, power management ICs, CMOS cameras, sensors, Li-ion batteries and displays. This is especially a concern for IoT edge devices which are intended to bring these different components together. 
 
\section{Discussion} \label{sec:discussion}

In this section, we put the results in perspective to illustrate other uses of the hardware framework and to account for additional considerations that can impact the final footprint, in the context of a massive deployment of IoT edge devices.

\subsection{Macroscopic carbon footprint analysis of massive IoT deployment} \label{sec:macroIoT}

The objective is to use our framework to estimate the global carbon footprint associated with the massive production of IoT edge devices over a ten years period (2018-2028). This problem is complex and uncertain by nature as it depends on many parameters such as the number of devices, the diversity of the applications and designs, the user behavior, the improvement of technology through time and its level of adoption. \\

\begin{figure*}[h!]
    \centering
    \includegraphics[width=1\textwidth]{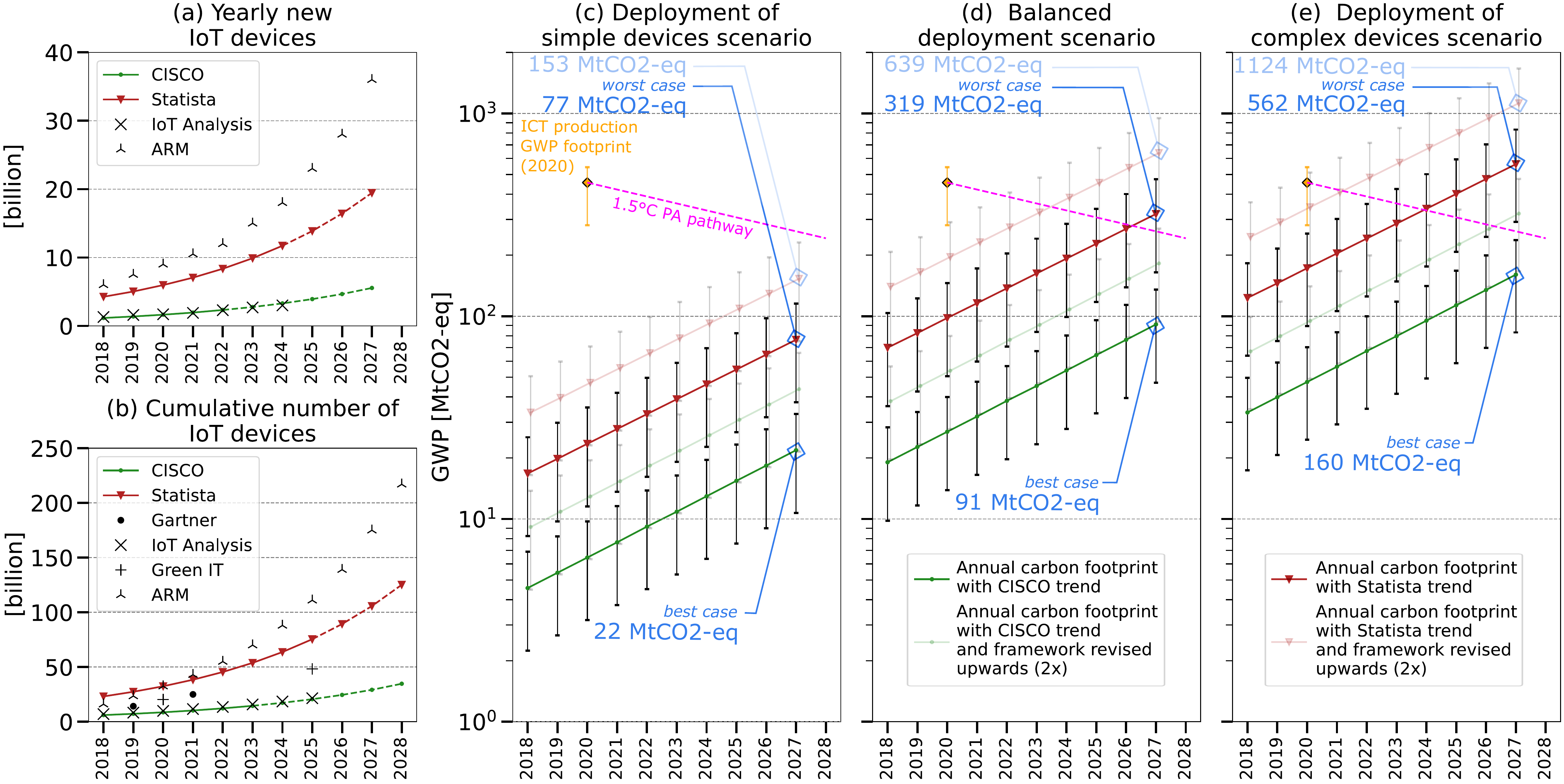}
    \caption{(a) Yearly deployment of new IoT devices computed based on the trends in (b) which represents the cumulative number of IoT devices, according to most popular market studies and predictions. Dashed lines are personal extrapolation. (c-e) Macroscopic analysis of the \textit{annual} carbon footprint generated by the production of IoT edge devices for different massive IoT deployment scenarios, based on (a). Shaded curves show the results if our framework is revised upwards by a factor $2 \times$ to account for the truncation error. (c) Scenario 1 \textit{deployment of simple devices} considers a majority of simple devices in the deployment i.e. $\alpha=90\%$ of light hardware profiles, (d) scenario 2 \textit{balanced deployment} considers a balanced mix of simple and complex devices i.e. $\alpha=50\%$, (e) scenario 3 \textit{deployment of complex devices} considers a majority of complex devices i.e. $\alpha=10\%$. }
    \label{fig:Macro_projections}
\end{figure*}

Regarding the number of devices, several trends exist \cite{CISCO, Statista_IoT, Gartner_2018, bordage-greenit2019, sparks2017route_ARM, IoTAnalytics} as shown in Fig. \ref{fig:Macro_projections}(a-b). Among them, we consider only the ones from CISCO \cite{CISCO} and Statista \cite{Statista_IoT} to model the massive deployment, as they seem to capture two main trends confirmed by the other sources. They do not include any traditional connected devices, e.g. computers, laptops, cell phones, smartphones, TVs or tablets. It is not always clearly defined if these trends account for gateways and if they consider all IoT edge devices. \par

To tackle the heterogeneous nature of IoT, hardware profiles are leveraged through different scenarios to picture these projections. Indeed, practical IoT deployment will be composed of many different hardware profiles and it would not be accurate to consider the same GWP for all nodes, as highlighted in Section \ref{sec:intepretation}. A simplified modeling approach is proposed in this study by considering the lowest and highest ranges of values given by our hardware framework. Additional details are available in the supplementary material. Three scenarios are proposed regarding the share of each hardware profile with respect to the number of devices deployed. In our analysis, the share of IoT edge devices with a light hardware profile is denoted as $\alpha$. A share $\alpha = 90\%$ is likely in the case of an IoT deployment based on small sensors with low functionality, similar to the occupancy sensor discussed previously. This is modeled in the first scenario called \textit{deployment of simple devices}. The second scenario called \textit{balanced deployment} models a deployment with an even share of simple and complex devices. The last scenario considers $\alpha = 10\%$ and is called \textit{deployment of complex devices}. \\

Results are presented in Fig. \ref{fig:Macro_projections} and show the importance to account at the same time for exponential trends and diversity in hardware profiles for IoT massive deployment. Shaded curves present the results if our framework is revised upwards by a factor $2\times$ to account for the truncation error mentioned in Section \ref{sec:comparison}. With a carbon footprint ranging from 22 to 153 MtCO2-eq/year, it is not likely that the massive production of simple devices will raise serious concerns from a carbon footprint point of view. However, if the majority of IoT edge devices turn to complex devices, the conclusion can change with a carbon footprint reaching 562 up to 1124 MtCO2-eq/year in 2027 worst-case scenarios. These scenarios will be favored if IoT edge devices embed increasing edge processing-memory abilities, wider user interfaces, bulky actuators and more advanced casings. For comparison, the annual carbon footprint of ICT production in 2020 lies between 281 and 543 MtCO2-eq \cite{freitag2021climate}. If the objectives of the Paris Agreement (PA) also apply to the production of ICT devices, it should follow a pathway as displayed in Fig. \ref{fig:Macro_projections}(c-e) i.e. a reduction of GHG emissions by $7.6\%$/year starting in 2020 to be consistent with the $1.5^{\circ} C$ target \cite{unepGAP-2019}. It clearly appears that the trends are conflicting : only the scenario \textit{deployment of simple devices} seems to stand within limits, even though this will likely generate concerns after 2030. It could be argued that this increasing carbon footprint will enable substantial reductions in other activity sectors (through the positive enabling effects mentioned in Section \ref{sec:intro}) and that only the final resulting trends matters. Both this question and the debate about allocation have to be more deeply investigated in future works but in any case, an important trade-off appears between the number of IoT edge devices and their average hardware profiles in order to guarantee that the carbon footprint for the production of IoT edge devices will remain a small share of current ICT footprint. \\

To the best of our knowledge, only two studies carried out a macroscopic analysis of the carbon footprint generated by massive IoT deployment \cite{malmodin2018energy, Das_IoT}. \par

On the one hand, \cite{malmodin2013future, malmodin2018energy} concluded that the total additional life cycle burden introduced by IoT devices was about 185-200 MtCO2-eq and that this will not impact significantly the carbon footprint of the ICT sector. However, they focused on a point estimate for a specific year (2015) which does not capture the exponential trends that characterize IoT massive deployment. They considered 27 billions connectivity modules for public displays, surveillance cameras, payment terminals, smart meters and wearables together with 500 billions tags and 1 billion small-cell wireless base stations.  Very few details about the modeling are available although it seems that the connectivity components were modeled based on the mobile connectivity devices for PCs \cite{malmodin2013future}. \par

On the other hand, \cite{Das_IoT} showed an exponential growing energy footprint for manufacturing that reach up to 2000 MtCO2-eq \footnote{\cite{freitag2021climate} did a conversion based on a global electricity mix of 0.63MtCO2eq/TWh but a primary energy factor has to be considered as \cite{Das_IoT} gives results in terms of primary energy. Neither \cite{freitag2021climate} nor \cite{Das_IoT} are peer-reviewed.} in 2025. However, they considered annual energy increase per device for hardware manufacturing on top of the exponential trend for IoT deployment, which is an assumption that we did not make in this study. Indeed, more advanced technologies need more energy to be manufactured \cite{garcia-2020} but at the same time they enable smaller die sizes and more integration \cite{bol2013green}. Nevertheless, emerging trends like edge computing \cite{adegbija2017microprocessor} could push towards more complex hardware at the edge. The evolution of the absolute carbon footprint over time is therefore uncertain \cite{BolDATE2021} and this highlights the need to carry out LCA for IoT edge devices early in the design phase.

\subsection{End-of-Life considerations}

In this study, the lifetime of IoT edge devices is not tackled. However in practical IoT deployment scenarios, real-life working conditions can introduce the need for replacement \cite{bonvoisin2012} due to aging of components leading to the accelerated death of nodes. This is likely to contribute to the already increasing number of abandoned IoT zombie devices due to poor life-cycle management \cite{soos2018iot}, especially in outdoor environments with corrosion and varying temperature-humidity conditions. Shortened effective lifetime for consumer IoT edge devices can also arise due to hardware, software or psychological obsolescence \cite{lehmann2010integrated}. Consequently, effective lifetime has to be differentiated from autonomy or expected lifetime. \\ 

In a recent paper \cite{ITU_WEEE}, attention is drawn on the additional contribution of IoT to wastes from electrical and electronic equipment (WEEE). They point out that WEEE from households and individual consumers is growing faster than infrastructure equipment waste mainly because of the growing number of individual devices, lifespans shortening and large amount of small hardware equipment used for the IoT. \textit{Small equipment} and \textit{small IT and telecommunication equipment} represent respectively 32\% and 9\% of the WEEE mass flow in 2019 \cite{GEM_2020}. As IoT edge devices will eventually become WEEE, end-of-life considerations must be addressed when discussing the impacts of IoT \cite{IoTSustBook}. \\

Assessing the real burden generated by the disposal of electronics is challenging. According to \cite{WEF_Reboot, GEM_2020}, about $80\%$ of WEEE are not documented nor collected for proper recycling. It acknowledges the fact that a wide majority ends up in developing countries where informal recycling and landfill occur, causing health and environmental concerns \cite{lehmann2010integrated}. The lack of representative data about environmental impacts explains partially why the vast majority of LCA discard this phase of the life cycle. Nevertheless, the end-of-life of electronics has usually a very low GWP contribution compared to production or use phases \cite{ercan2016}. This does not necessarily hold for impacts in other categories such as freshwater, terrestrial and human eco-toxicity.

\subsection{Rebound effects considerations}

Rebound effect, also called \textit{Jevons paradox}, describes the fact that technological efficiency gains do not \textit{de facto} lead to absolute resources or energy savings. The dynamics behind rebound effects is complex and hard to quantify \cite{coroamua2020methodology}, especially for indirect and structural effects which occur dynamically across space and time at a society scale \cite{hilty2015ict, gossart2015rebound}. Nevertheless, the conditions that encourage the development of rebound effects are relatively well understood \cite{gossart2015rebound, wallenborn2018rebounds, binswanger2001technological, schneider2009importance}, e.g. policies and designs focusing only on energy-efficiency; overlooking the user behavior regarding effective lifetime and effective use of the devices; end users with high incomes; high inter-connectivity; access to high quality energy such as electricity; time-savings and money-savings motivations. \\

Several of these conditions are met for IoT, which creates a fertile ground for rebound effects. For instance, if IoT devices become cheaper and cheaper, people will tend to buy more and this could result in an absolute increase both in terms of hardware and energy consumption \cite{hittinger2019} (even without accounting for obsolescence mechanisms). Rebound effects are very difficult to predict for emerging technologies, especially in the long-term \cite{hittinger2019}. Indeed, new usages and services will emerge with the massive introduction of IoT. As stated by Hilty for the ICT in general, an efficiency strategy must always be accompanied by a sufficiency strategy \cite{hilty2011information}, which is also supported by \cite{BolDATE2021}. Sobriety must be applied during the IoT deployment if one wants to avoid rebound effects to backfire the savings it can enable. Consequently, IoT designers are called to consider the environmental impacts generated by their devices over the whole life-cycle, both at the device and system levels. \\
\section{Conclusion and future perspectives}

The Internet-of-Things is expected to boom in the upcoming years, supported by the faith in the promising positive enabling effects it could bring. However, potential benefits of a massive IoT deployment will come together with undeniable environmental burdens which are usually overlooked. Indeed, life-cycle assessments for IoT devices are really scarce, both in academia and industry. \\

To take a step towards a better consideration of the environmental impacts generated by IoT edge devices, this paper proposes a framework to quantify the carbon footprint due to the production and transport of such devices. The framework is based on state-of-the-art existing tools and databases and it also helps to cope with the wide variety of IoT devices by introducing hardware profiles. We showed that the carbon footprint of IoT edge devices with different hardware profiles can vary up to a factor $158\times$. We applied the framework on four IoT use cases and we demonstrated that it can be used to streamline their carbon footprints, which significantly reduces the uncertainty down to a factor $3-5\times$. We then estimated the carbon footprint related to the production of IoT edge devices for massive deployment over a 10-year period. Depending on the scenario, it could range from 22 to 1124 MtCO2-eq/year in 2027, which points out the critical need to consider both the exponential trend for the number of devices and the heterogeneous nature of IoT. This raises concerns regarding consumer electronics as it might flood the market with significantly more complex IoT edge devices. \\

Further work is needed to (1) take into account the remaining phases of the life cycle, (2) include the associated use of networks and infrastructures and (3) to take more indicators into consideration, e.g. cumulative energy, water consumption and toxicity. Indeed, GWP scratches only a small fraction of environmental impacts and sustainability challenges. There is a critical need for LCA in the context of IoT to guide decision-making processes towards the potential benefits of IoT without worsening the current environmental situation.
\section*{Acknowledgments}

This work was supported by the Fonds européen de développement régional (FEDER) and the Wallonia within the Wallonie-2020.EU program.

\section*{Supplementary material}

The Supporting Information includes additional details about the carbon footprint of the framework; modeling limitations; the mapping between use cases and their corresponding hardware profile; memory density trends analysis; the Google Home mini teardown; and the macroscopic analysis.

\bibliography{bibliography} 
\bibliographystyle{apalike}

\clearpage

\begin{center}
\vspace{0.5cm}

\begin{center}
    \includegraphics[width=0.6\textwidth]{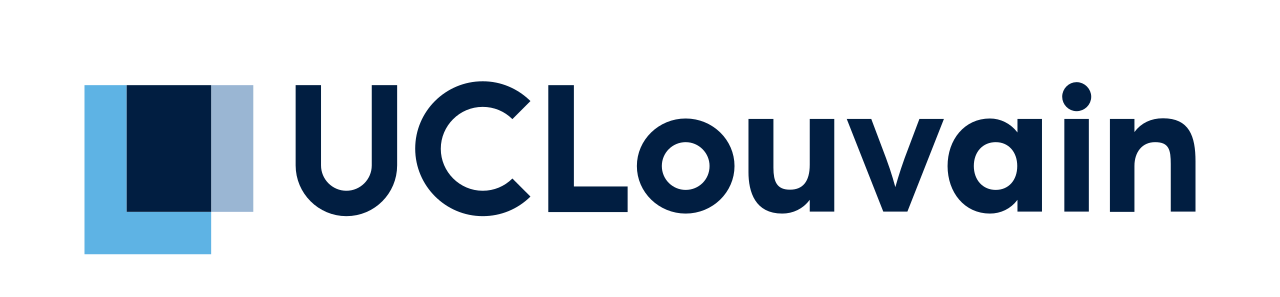}
\end{center}

\vspace{1cm}

\rule{\linewidth}{0.5mm} \\[0.1cm]
\centering
\textbf{\LARGE{Supplementary material \\[1cm] Assessing the embodied carbon footprint of IoT edge devices with a bottom-up life-cycle approach}}\\[0.5cm]
\large{Thibault Pirson$^{a,*}$ and David Bol$^a$} \\
\flushleft
\small{$^a$\textit{Université catholique de Louvain, ICTEAM/ECS, Louvain-la-Neuve, Belgium}} \\
\small{$^*$\textit{Corresponding author at:} Bâtiment Maxwell a.193, Place du Levant, 3 (L5.03.02) 1348 Louvain-la-Neuve, Belgium.} \\
\small{\textit{Email address:} thibault.pirson@uclouvain.be}\\
\rule{\linewidth}{0.5mm} \\[1cm]
\centering

\vfill
\hrulefill\par
\today\\
\end{center}


\clearpage
\section{Carbon footprint of the framework}

In order to ease re-uptake of the framework, the carbon footprint of each \textit{functional block} and \textit{hardware specification level} are given in Table \ref{tab:framework_cf}. The total row is the sum for a specific \textit{hardware specification level}, across all \textit{functional blocks}. The sum of minimum and maximum values obtained in the framework with respect to \textit{functional blocks} are respectively 0.3 kgCO2-eq and 47.41 kgCO2-eq. Hence, the ratio of 158$\times$ between extrema mentioned in the manuscript.

\begin{table}[h!]
    \caption{Carbon footprint for each \textit{functional block} and \textit{hardware specification level} in the framework. Lower, typical and upper values are given in kgCO2-eq, according to Table 1 in the manuscript. }
    \label{tab:framework_cf}
    
    \centering
    \begin{tabular}{l cccccccccccc}
    \toprule
        & \multicolumn{12}{c}{\textbf{Hardware Specification Level (HSL)}} \\
        \cmidrule{2-13}
        \textbf{Block} & \multicolumn{3}{c}{HSL-0} & \multicolumn{3}{c}{HSL-1} & \multicolumn{3}{c}{HSL-2} & \multicolumn{3}{c}{HSL-3} \\ 
        \cmidrule{3-3} \cmidrule{6-6} \cmidrule{9-9} \cmidrule{12-12}
        \textbf{Function} & low & typical & up & low & typical & up & low & typical & up & low & typical & up \\ 
        \midrule
        
        Actuators & 0.00 & 0.00 & 0.00 & 0.03 & 0.06 & 0.12 & 0.17 & 0.33 & 0.66 & 1.03 & 4.12 & 6.19 \\ 
        Casing & 0.00 & 0.00 & 0.00 & 0.04 & 0.27 & 0.63 & 0.88 & 2.13 & 3.14 & 3.06 & 4.26 & 5.93 \\ 
        Connectivity & 0.00 & 0.00 & 0.00 & 0.08 & 0.17 & 0.34 & 0.53 & 0.82 & 1.26 & 1.21 & 1.36 & 1.63 \\ 
        Memory & 0.00 & 0.00 & 0.00 & 0.05 & 0.18 & 0.74 & 0.82 & 1.91 & 3.63 & 1.88 & 3.73 & 7.27 \\ 
        Others & 0.06 & 0.11 & 0.14 & 0.14 & 0.22 & 0.31 & 0.28 & 0.41 & 0.54 & 0.49 & 0.64 & 0.85 \\ 
        PCB & 0.13 & 0.16 & 0.24 & 0.24 & 0.57 & 0.81 & 1.12 & 1.60 & 3.20 & 2.56 & 3.84 & 4.80 \\ 
        Power Supply & 0.18 & 0.52 & 0.66 & 0.02 & 0.18 & 0.38 & 0.25 & 1.36 & 2.71 & 0.37 & 1.72 & 3.32 \\ 
        Processing & 0.08 & 0.17 & 0.29 & 0.66 & 1.05 & 1.60 & 1.58 & 1.94 & 2.52 & 2.31 & 3.13 & 3.98 \\ 
        Security & 0.00 & 0.00 & 0.00 & 0.01 & 0.02 & 0.03 & N/A & N/A & N/A & N/A & N/A & N/A \\ 
        Sensing & 0.00 & 0.00 & 0.00 & 0.01 & 0.02 & 0.04 & 0.00 & 0.03 & 0.04 & 0.19 & 0.77 & 1.47 \\ 
        Transport & 0.00 & 0.00 & 0.00 & 0.18 & 0.40 & 1.35 & 1.07 & 2.62 & 4.05 & 3.34 & 6.30 & 9.34 \\ 
        UI & 0.00 & 0.00 & 0.00 & 0.03 & 0.06 & 0.12 & 0.19 & 0.75 & 2.60 & 0.17 & 0.61 & 2.01 \\ 
        \toprule
        Total & 0.46 & 0.96 & 1.33 & 1.49 & 3.20 & 6.47 & 6.89 & 13.88 & 24.36 & 16.62 & 30.47 & 46.80 \\ 
        \toprule
    \end{tabular}
    
\end{table}

\vspace{-0.7cm}


\section{Modeling assumptions: limitations}

This section aims at highlighting known limitations of the framework both for the sake of completeness and for potential future works. \\

No recycled material has been explicitly introduced when modeling the IoT edge devices in GaBi: all materials are virgin materials. When an entry of the database is scaled in GaBi, acquisition processes are scaled accordingly. Regarding the cut-off embedded in GaBi databases for specific processes already modeled, please refer to their documentation. Transportation within the processes is taken into account in the databases but exact details are not available in the documentation. Electricity mixes are location-based and are also taken into account in the databases. For ICs, thermal pad area is used as a proxy for die size when no details are given. All silicon area in the framework are given in equivalent silicon die area.

\subsection{Actuators}

\textit{Limitations:} the very scarce information regarding material declaration for the actuators complicates proper modeling. The presence of rare earths materials in the magnet needs more in-depth modeling and is likely to increase the footprint \cite{marx2018comparative}. The modeling of relays could be extended to other types (e.g. electromechanical).

\subsection{Casing}

\textit{Limitations:} casings with an important share of metals, heterogeneous materials, and processes need more in-depth modeling. The framework did not model hardware sunk in resin for outdoor nodes (waterproof), which is a common practice. Note that this significantly complicates the recycling of nodes.

\subsection{Connectivity}

\textit{Limitations:} die inspections and statistical analysis are needed for communication ICs because literature and LCA databases are very scarce for communication devices \cite{Proske_FF3}. To the best of our knowledge, no data base tackles this issue. A better modeling of the wide range of antennas would also be of interest.

\subsection{Memory}

\begin{figure}[h!]
    \centering
    \includegraphics[width=0.7\textwidth]{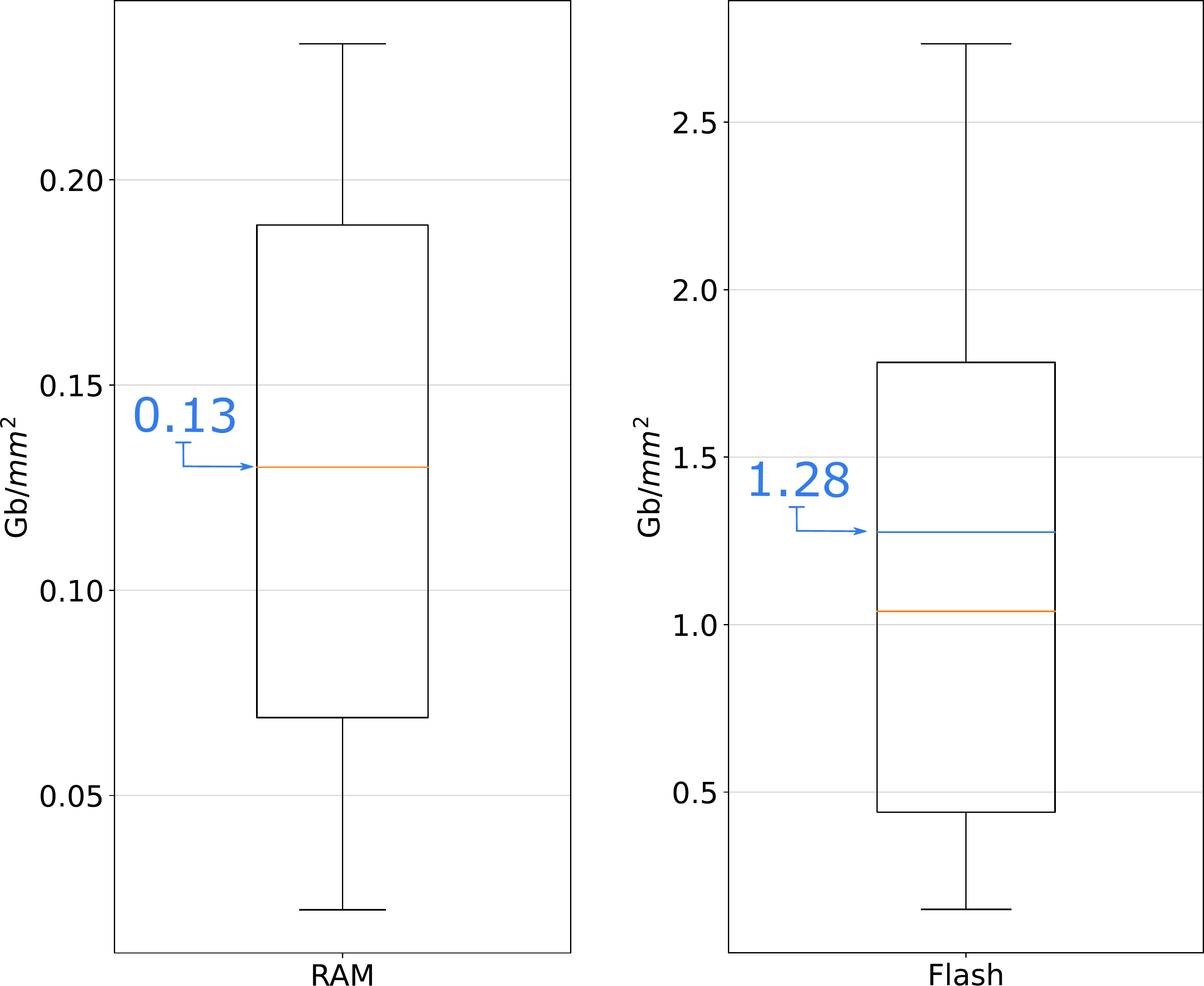}
    \caption{Boxplots for the memory trends over more than 10 years, RAM and Flash.}
    \label{fig:memory_trends}
\end{figure}

\textit{Limitations:} the comparison of silicon area obtained from the thermal pad estimation and memory density trends (Gb/$mm^2$) over the last decades reveals a potential overshoot when thermal pad area is used as a proxy for die size. For this reason, memory density has been considered when no other information was available: 0.13 Gb/$mm^2$ has been used for (S/D)RAM and 1.28 Gb/$mm^2$ for Flash. Several sources have been analyzed with trends over more than 10 years (2010-2022). All details and sources are given in the file memory\_density\_trends.xlsx. Figure \ref{fig:memory_trends} shows two boxplots for the RAM and Flash densities across the different sources considered. These densities seem coherent with recent work \cite{Schischke2021-EUreport} which evaluates about 0.22 Gb/$mm^2$ for DRAM and from 1.28 to 2.56 Gb/$mm^2$ for Flash. Nevertheless, they focus on smartphones only. \\

The die-to-package ratios given in GaBi documentation suggest that stack dies are not taken into account nor memory capacity. This is problematic as it is more common for memory ICs to find stack dies in a single package. Stacked dies are usually thinner than non-stacked dies but total silicon die area is considerably increased, so are the environmental impacts. Die inspections and statistical analysis would be useful to better evaluate silicon die area in memory packages. Another limitation is that 45/57 nm are quite old technologies for memories but these are the only ones available in the GaBi Electronics Extension database.

\subsection{Printed circuit board (PCB)}

\textit{Limitations:} 4 to 8 routing metal layers are considered in this framework but for highly integrated designs, this could go above 10 layers which is not directly captured here. A reasonable conversion can be done by multiplying area and number of layers. Regarding the solder paste, it is a rough approximation.

\subsection{Power supply}

The emission factor for small alkaline batteries are similar between the Base Carbone and GaBi. More specifically, Base Carbone gives 0.0653 kgCO2-eq/unit and 0.137 kgCO2e/unit for AAA and AA alkaline batteries while GaBi gives 0.09 kgCO2-eq and  0.189 kgCO2-eq/unit, respectively. \\

\textit{Limitations:} energy harvesting is not covered by this framework. Even though it is increasingly used for industrial IoT applications, it is less the case for consumer IoT edge devices. Energy harvesting has the potential to extend the lifetime of the IoT edge devices and to decrease the need of maintenance but this comes at the expense of some additional hardware components e.g. energy harvesters, supercapacitors and power management ICs. This trade-off should be considered in future works in regards of environmental impacts. 

\subsection{Processing}

\textit{Limitations:} die inspections and statisical analysis of several IoT processing ICs would help to better estimate the effective silicon die area and content of ICs. 

\subsection{Security}

\textit{Limitations:} only small external security ICs are considered.

\subsection{Sensing}

\textit{Limitations:} electro-chemical and non-CMOS technologies are not captured by this framework.

\subsection{Transport}

\textit{Limitations:} this is a simplified view of real world logistics which likely underestimates actual distances and the diversity in the modes of transportation.

\subsection{User interfaces (UI)}

\textit{Limitations:} large displays are not captured in this framework. Regarding GWP impacts, \cite{louis2020sources} points out that the impact of upstream suppliers and the production of input material is not covered in \cite{AUO}. This might explain at least partially why the energy per $m^2$ of the Taiwanese display manufacturer is about one order of magnitude lower than in \cite{ercan2016} i.e. $0.008 kWh/cm^2$ compared to $0.1 kWh/cm^2$, respectively.

\subsection{Others}

\textit{Limitations:} connectors are not covered (cut-off).


\section{Use cases : additional details}

This section gives additional details about the use cases considered in the manuscript. As highlighted in both ETSI-ITU standards and literature \cite{ercan2016, Proske_FF3}, assessing the life-cycle impacts of ICT products is a very complex task. Indeed, each device is made up from different electronic components and very few details regarding environmental impacts are publicly available in the supply chain. The exact quantification of direct impacts associated with such devices is therefore out of reach. This complexity is further enhanced for IoT due to the wide variety of applications and designs.

\subsection{Relative carbon footprint contributions}

\begin{figure}[h!]
    \centering
    \includegraphics[width=1\textwidth]{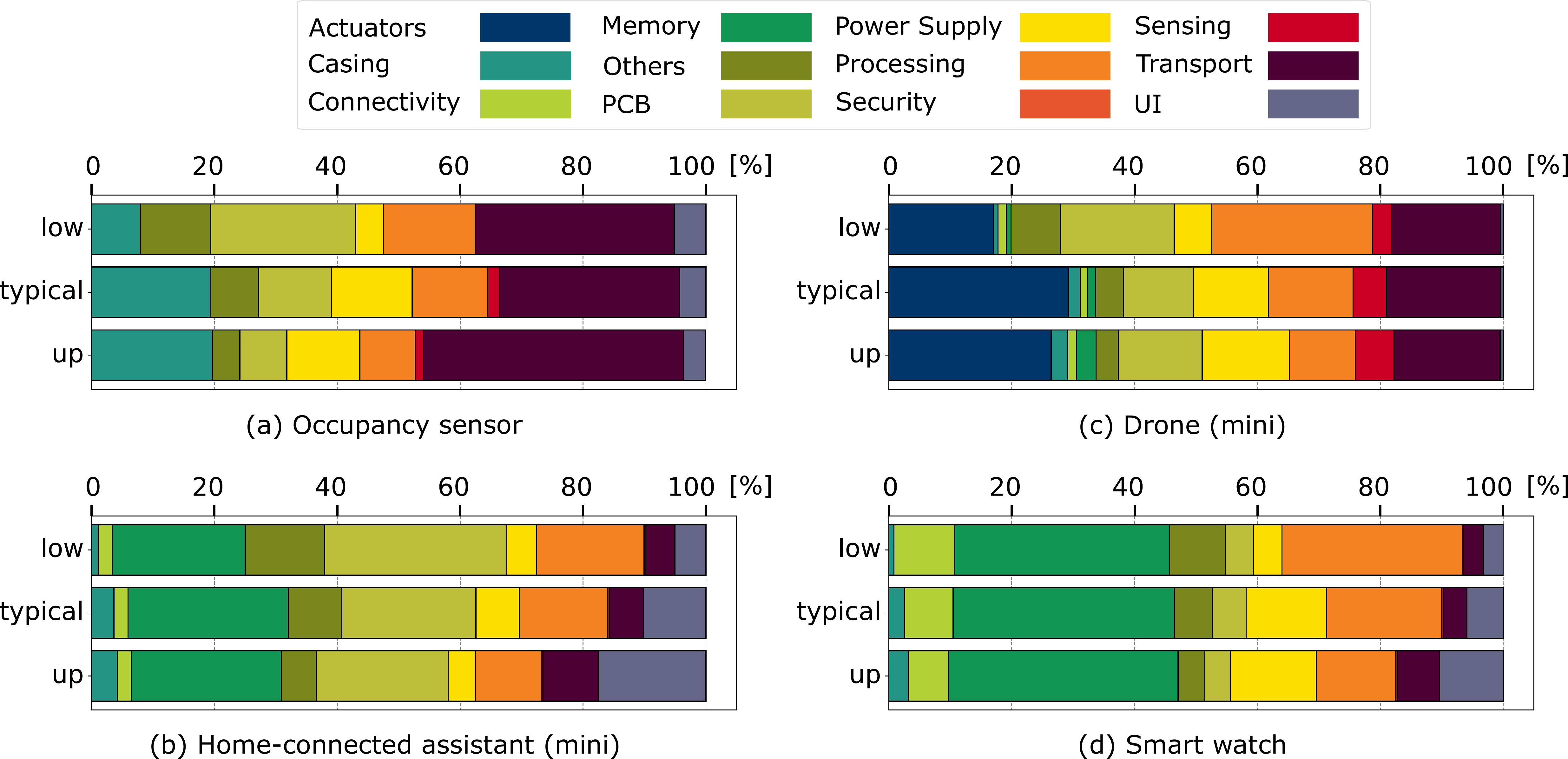}
    \caption{Relative contribution of each \textit{functional block} to the total carbon footprint of each use case.}
    \label{fig:all_results_proc}
\end{figure}

Figure \ref{fig:all_results_proc} shows the relative contribution of each \textit{functional block} to the total carbon footprint of each use case. This supports the interpretation and discussion detailed in the manuscript. A better estimate of the carbon footprint for a given hardware profile can be obtained by combining only the parameters that best model the hardware of the IoT edge devices: this can be done by using Table \ref{tab:framework_cf}. 

\newpage 
\subsection{Mapping between use cases and corresponding hardware profiles}

The hardware profiles given in the article are based on a mapping resulting of the analysis of the electronic components available for each use case. All details are given in the file mapping\_usecases.xlsx. It was more difficult to have details for on-line teardown reports, which explains why several information are missing in the Excel file. This points out again the important limitations existing when applying LCA to IoT edge devices and more generally to electronic devices. 

\subsection{Google Home mini teardown}

The whole teardown of the Google Home mini was performed in our labs. First, we carefully dismantled the device in order to preserve all materials and parts. Then, we removed all electromagnetic shields to access ICs. All ICs have been desoldered using a rework station. We then identified ICs when it was possible, based on the information available on the packages. Datasheets have been gathered when possible and all ICs were measured and weighted using a Statorius balance (d = 0.1 mg). \\

In order to obtained the exact die size contained in each ICs, we used a chemical die inspection method to disintegrate epoxy packages. Several inspection methods could be used \cite{aryan2018_die_inspection} but they can be expensive and demanding in terms of infrastructures, e.g. X-ray and micro X-ray fluorescence spectrometry (micro-XRF). To simplify the process, we used a $98\%$ sulfuric acid solution ($H_2SO_4$) while heating up to $240^\circ$C under a gas extractor, as proposed by DeusExSilicium \cite{deusexsilicium}. It took from 3 to 15 minutes for packages to be completely disaggregated. We then cleaned dies with an ultrasonic water bath and we weighted the naked dies. Finally we used an optic microscope with a camera to inspect dies and measure the actual die size precisely. Unfortunately, it was not possible to identify the technology nodes due to the important number of dense routing layers above transistors. We checked for stacked dies but we did not found any. \\

All results are given in the tab \textit{Google Home mini - teardown} within the file mapping\_usecases.xlsx. Note that all this work could have been avoided if the necessary information were available in the datasheets. 


\section{Macroscopic carbon footprint analysis of massive IoT deployment}

\subsection{IoT massive deployment trends}

Several trends exist regarding the evolution of the number of IoT devices in the upcoming years \cite{CISCO, Statista_IoT, Gartner_2018, bordage-greenit2019, sparks2017route_ARM, IoTAnalytics}. In this study, we focused on the ones from CISCO \cite{CISCO} and Statista \cite{Statista_IoT} to model the massive deployment, as they seem to capture two main trends confirmed by the other sources. Extrapolation of the trends over a couple of years was done such that they all fit the same timeframe, i.e. 2018-2028. While some trends are given in a cumulative number, others are given on a yearly basis. Because the carbon footprint of ICT is usually expressed in terms of MtCO2-eq/year, we converted cumulative trends to differential ones, when sufficient data were available. Tables \ref{tab:cumul} and \ref{tab:diff} detail the trends for the different sources. According to Statista, this data does not include any computers, laptops, fixed line phones, cell phones or tablets. Active devices or gateways that concentrate the end-sensors are considered, but not every sensor-actuator.

\begin{table}[h!]
    \caption{Cumulative number of IoT devices (billion), according to most popular market studies and predictions.}
    
    \label{tab:cumul}
    
    \centering
    \begin{tabular}{ l r r r r r r r r r r r }
     \toprule
         & 2018 & 2019 & 2020 & 2021 & 2022 & 2023 & 2024 & 2025 & 2026 & 2027 & 2028 \\  
         \midrule
        CISCO & 6.1 & 7.26 & 8.64 & 10.28 & 12.23 & 14.56 & 17.32$^\dagger$ & 20.61$^\dagger$ & 24.53$^\dagger$ & 29.19$^\dagger$ & 34.74$^\dagger$ \\  
        Statista & 23.14 & 27.4 & 32.43 & 38.4 & 45.46 & 53.82 & 63.72 & 75.44 & 89.31$^\dagger$ & 105.74$^\dagger$ & 125.19$^\dagger$ \\  
        Gartner & N/A & 14.5 & N/A & 25 & N/A & N/A & N/A & N/A & N/A & N/A & N/A \\  
        IoT Analytics & 7 & 8.3 & 9.9 & 11.6 & 13.5 & 15.8 & 18.5 & 21.5 & N/A & N/A & N/A \\  
        GreenIT & N/A & N/A & 20.315 & N/A & N/A & N/A & N/A & 48.272 & N/A & N/A & N/A \\
        \toprule
    \end{tabular}
    \vspace{-0.4cm}
    \flushleft{$^\dagger$personal extrapolation based on the previous trend.\\}
    
\end{table}

\begin{table}[h!]
\caption{Yearly deployment of new IoT devices (billion) computed based on the trends in Table \ref{tab:cumul}.}
    
    \label{tab:diff}
    \centering
    \begin{tabular}{ l r r r r r r r r r r r }
     \toprule
         & 2018 & 2019 & 2020 & 2021 & 2022 & 2023 & 2024 & 2025 & 2026 & 2027 & 2028 \\  
         \midrule
        CISCO & 1.16 & 1.38 & 1.64 & 1.95 & 2.33 & 2.76 & 3.29$^\dagger$ & 3.92$^\dagger$ & 4.66$^\dagger$ & 5.55$^\dagger$ & N/A \\  
        Statista& 4.26 & 5.03 & 5.97 & 7.06 & 8.36 & 9.9 & 11.72  & 13.87$^\dagger$ & 16.43$^\dagger$ & 19.45$^\dagger$ & N/A \\  
        IoT Analytics & 1.3 & 1.6 & 1.7 & 1.9 & 2.3 & 2.7 & 3 & N/A & N/A & N/A & N/A \\  
        ARM & 6 & 7.5 & 9 & 10.5 & 12 & 15 & 18 & 23 & 28 & 36 & 42 \\  
        \toprule
    \end{tabular}
    
    \vspace{-0.4cm}
    \flushleft{\qquad $^\dagger$personal extrapolation based on the previous trend.\\
    }
\end{table}

\subsection{Carbon footprint projections by scenarios}

A simplified modeling approach was proposed in this study in order to estimate the global carbon footprint associated with the massive production of IoT edge devices over a ten years period (2018-2028). \\

To account for the heterogeneity of IoT edge devices, we leveraged our framework by creating two hardware profiles for simple and complex devices. Simple devices are defined by the lowest value obtained in the framework and the carbon footprint resulting of the sum of all \textit{HSL-0}. This yields a range $D_s =$ 0.3-1.3 kgCO2-eq, as shown in Table \ref{tab:framework_cf}. Similarly, complex devices are defined by the highest value obtained in the framework and the carbon footprint resulting of the sum of all \textit{HSL-3}. This yields a range $D_c =$ 16.6-47.4 kgCO2-eq, as shown in Table \ref{tab:framework_cf}. \\

Then, we proposed three simple scenarios based on the share of IoT edge devices with a light hardware profile, denoted as $\alpha$ in our analysis ($0<\alpha<1$). Scenario 1 \textit{deployment of simple devices} considers a majority of simple devices in the deployment i.e. $\alpha=90\%$ of light hardware profiles. Scenario 2 \textit{balanced deployment} considers a balanced mix of simple and complex devices i.e. $\alpha=50\%$. Scenario 3 \textit{deployment of complex devices} considers a majority of complex devices i.e. $\alpha=10\%$. The annual carbon footprint $F_y$ for a given year $y$ is therefore given by 

\begin{equation}
F_y = N_y \big( \alpha D_s + (1-\alpha) D_c \big)
\end{equation}

where $N_y$ is the number of IoT devices deployed during the year $y$. Moreover, as we pointed out that the results obtained by our framework tend to undershoot the carbon footprint when compared to results from existing environmental reports due to the truncation error, we propose to compensate for this through the parameter $\psi$. Therefore, the annual carbon footprint with the framework revised upwards ($F^R_y$) turns to $F^R_y = \psi F_y$. A value of $\psi = 2$ was considered in this study to account . Table \ref{tab:projections} gives the resulting values of $F_y$ and $F^R_y$ for $y$ ranging from 2018 to 2027.

\begin{table}[h!]

    \caption{Macroscopic analysis of the annual carbon footprint generated by the production of IoT edge devices for different massive IoT deployment scenarios, in MtCO2-eq/year. SC stands for scenario. Bold values are the ones highlighted in the manuscript.}

    \label{tab:projections}

    \centering
    \begin{tabular}{l l r r r r r r r r r r r }
    \toprule
         &  &  & 2018 & 2019 & 2020 & 2021 & 2022 & 2023 & 2024 & 2025 & 2026 & 2027 \\ \midrule
        SC-1 & CISCO$^\blacktriangle$  & low & 2.2 & 2.7 & 3.2 & 3.8 & 4.5 & 5.3 & 6.4 & 7.6 & 9.0 & 10.7 \\  
         &  & typical & 4.6 & 5.4 & 6.5 & 7.7 & 9.2 & 10.9 & 12.9 & 15.4 & 18.3 & \textbf{21.8} \\  
         &  & up & 6.9 & 8.2 & 9.7 & 11.6 & 13.8 & 16.4 & 19.5 & 23.3 & 27.7 & 33.0 \\  
         & CISCO$^\vartriangle$ & low & 4.5 & 5.3 & 6.3 & 7.5 & 9.0 & 10.7 & 12.7 & 15.1 & 18.0 & 21.4 \\  
         &  & typical & 9.1 & 10.9 & 12.9 & 15.3 & 18.3 & 21.7 & 25.9 & 30.9 & 36.7 & 43.7 \\  
         &  & up & 13.8 & 16.4 & 19.5 & 23.2 & 27.7 & 32.8 & 39.1 & 46.6 & 55.3 & 65.9 \\  
         & Statista$^\blacktriangle$  & low & 8.2 & 9.7 & 11.5 & 13.6 & 16.2 & 19.1 & 22.6 & 26.8 & 31.7 & 37.6 \\  
         &  & typical & 16.8 & 19.8 & 23.5 & 27.8 & 32.9 & 39.0 & 46.1 & 54.6 & 64.7 & \textbf{76.5} \\  
         &  & up & 25.3 & 29.9 & 35.4 & 41.9 & 49.6 & 58.8 & 69.6 & 82.4 & 97.6 & 115.5 \\  
         & Statista$^\vartriangle$ & low & 16.5 & 19.4 & 23.1 & 27.3 & 32.3 & 38.3 & 45.3 & 53.6 & 63.5 & 75.2 \\  
         &  & typical & 33.5 & 39.6 & 47.0 & 55.6 & 65.8 & 77.9 & 92.2 & 109.2 & 129.3 & \textbf{153.1} \\  
         &  & up & 50.6 & 59.7 & 70.9 & 83.8 & 99.3 & 117.6 & 139.2 & 164.7 & 195.1 & 231.0 \\ 
         
         \midrule
         
        SC-2 & CISCO$^\blacktriangle$  & low & 9.8 & 11.7 & 13.9 & 16.5 & 19.7 & 23.3 & 27.8 & 33.2 & 39.4 & 47.0 \\  
         &  & typical & 19.0 & 22.7 & 26.9 & 32.0 & 38.2 & 45.3 & 54.0 & 64.3 & 76.5 & \textbf{91.1} \\  
         &  & up & 28.3 & 33.6 & 40.0 & 47.5 & 56.8 & 67.3 & 80.2 & 95.5 & 113.6 & 135.3 \\  
         & CISCO$^\vartriangle$ & low & 19.6 & 23.3 & 27.7 & 33.0 & 39.4 & 46.7 & 55.7 & 66.3 & 78.8 & 93.9 \\  
         &  & typical & 38.1 & 45.3 & 53.8 & 64.0 & 76.5 & 90.6 & 108.0 & 128.7 & 153.0 & 182.2 \\  
         &  & up & 56.5 & 67.3 & 79.9 & 95.0 & 113.6 & 134.5 & 160.4 & 191.1 & 227.1 & 270.5 \\  
         & Statista$^\blacktriangle$  & low & 36.0 & 42.6 & 50.5 & 59.7 & 70.7 & 83.8 & 99.2 & 117.3 & 139.0 & 164.5 \\  
         &  & typical & 69.9 & 82.6 & 98.0 & 115.9 & 137.2 & 162.5 & 192.4 & 227.7 & 269.7 & \textbf{319.3} \\  
         &  & up & 103.8 & 122.6 & 145.5 & 172.1 & 203.7 & 241.3 & 285.6 & 338.0 & 400.4 & 474.0 \\  
         & Statista$^\vartriangle$ & low & 72.1 & 85.1 & 101.0 & 119.5 & 141.5 & 167.5 & 198.3 & 234.7 & 278.0 & 329.1 \\  
         &  & typical & 139.9 & 165.1 & 196.0 & 231.8 & 274.5 & 325.0 & 384.8 & 455.4 & 539.4 & \textbf{638.5} \\  
         &  & up & 207.6 & 245.2 & 291.0 & 344.1 & 407.5 & 482.5 & 571.2 & 676.0 & 800.8 & 948.0 \\  
         
         \midrule
         
        SC-3 & CISCO$^\blacktriangle$  & low & 17.4 & 20.7 & 24.6 & 29.2 & 34.9 & 41.4 & 49.3 & 58.8 & 69.8 & 83.2 \\  
         &  & typical & 33.5 & 39.9 & 47.4 & 56.3 & 67.3 & 79.8 & 95.1 & 113.3 & 134.7 & \textbf{160.4} \\  
         &  & up & 49.7 & 59.1 & 70.2 & 83.5 & 99.7 & 118.1 & 140.8 & 167.8 & 199.5 & 237.6 \\  
         & CISCO$^\vartriangle$ & low & 34.8 & 41.4 & 49.2 & 58.5 & 69.8 & 82.7 & 98.6 & 117.5 & 139.7 & 166.4 \\  
         &  & typical & 67.0 & 79.8 & 94.8 & 112.7 & 134.7 & 159.5 & 190.1 & 226.5 & 269.3 & 320.7 \\  
         &  & up & 99.3 & 118.1 & 140.4 & 166.9 & 199.5 & 236.3 & 281.6 & 335.6 & 398.9 & 475.1 \\  
         & Statista$^\blacktriangle$  & low & 63.8 & 75.4 & 89.5 & 105.8 & 125.3 & 148.4 & 175.7 & 207.9 & 246.3 & 291.5 \\  
         &  & typical & 123.1 & 145.3 & 172.5 & 204.0 & 241.6 & 286.1 & 338.6 & 400.8 & 474.7 & \textbf{562.0} \\  
         &  & up & 182.3 & 215.3 & 255.5 & 302.2 & 357.8 & 423.7 & 501.6 & 593.7 & 703.2 & 832.5 \\  
         & Statista$^\vartriangle$ & low & 127.7 & 150.8 & 179.0 & 211.6 & 250.6 & 296.8 & 351.3 & 415.8 & 492.5 & 583.0 \\  
         &  & typical & 246.2 & 290.7 & 345.0 & 408.0 & 483.1 & 572.1 & 677.3 & 801.5 & 949.5 & \textbf{1124.0} \\  
         &  & up & 364.7 & 430.6 & 511.1 & 604.4 & 715.6 & 847.5 & 1003.3 & 1187.3 & 1406.5 & 1665.0 \\  
         \toprule
    \end{tabular}
    
\vspace{-0.3cm}
\flushleft{
$^\blacktriangle$ Framework baseline ($\psi = 1$). $^\vartriangle$ Framework revised upwards ($\psi = 2$)
}

\end{table}



\end{document}